\documentclass[fleqn,usenatbib]{mnras}

\usepackage{graphicx}	% Including figure files
\usepackage{amsmath}	% Advanced maths commands
\usepackage[dvipsnames]{xcolor}  
\usepackage{amssymb}	% Extra maths symbols
\usepackage{newtxtext,newtxmath}    % PUT THIS ONE 
\usepackage{booktabs}
\usepackage{siunitx}
\usepackage{placeins}
\usepackage{ulem}
\usepackage{soul}
\usepackage{enumitem}
\usepackage{orcidlink}

\usepackage{macros_vdbosch} % my personal macros

\title[Transient Lensing of SIDM Haloes]
{Through Ups and Downs: the Transient Lensing Behaviour of SIDM halos undergoing Core Collapse}

\author[F.~C.~van den Bosch and Y.~M.~Tokayer]{
Frank C. van den Bosch$^1$\thanks{E-mail: frank.vandenbosch@yale.edu}\orcidlink{0000-0003-3236-2068}
and
Yarone M. Tokayer$^2$\orcidlink{0000-0002-0430-5798}
\vspace*{8pt}
\\
   $^1$Department of Astronomy, Yale University, PO. Box 208101, New Haven, CT 06520-8101\\
   $^2$Department of Physics, Yale University, PO. Box 208120, New Haven, CT 06520-8120
}

\date{Draft: \today}
\pubyear{2026}

\begin{document}

\label{firstpage}
\pagerange{\pageref{firstpage}--\pageref{lastpage}}
\maketitle

\begin{abstract}
  Self-interacting dark matter (SIDM) haloes can become extremely dense following gravothermal core collapse, motivating their use as explanations for unusually concentrated gravitational lenses.  Using an emulator calibrated on gravothermal fluid calculations, we investigate whether increasing central density necessarily strengthens lensing. We follow initially Navarro–Frenk–White (NFW) haloes through core formation and collapse deep into the short-mean-free-path regime, and compute finite-source magnifications, critical curves, and source-plane lensing cross sections for isolated spherical haloes with elastic, velocity-independent self-interactions, including external shear in selected calculations. Lensing strength evolves non-monotonically: it initially decreases during core formation, rises during early collapse, and generally declines at later stages despite continued growth of the central density. This decline accompanies the emergence of a dense inner core whose mass and radius decrease within a more extended, lower-density core. Time-averaged enhancement relative to NFW haloes depends on the lensing statistic, halo concentration, and interaction cross section. Generalized NFW haloes with central cusps $\rho \propto r^{-\gamma}$ with $\gamma \geq 2$ always have larger lensing efficacies than core-collapsed SIDM haloes of equal mass and are therefore poor proxies, including in terms of the demographics (size and radial thickness) of arcs and Einstein rings. The core collapse stage at which SIDM haloes are optimal lenses coincides with the epoch at which the characteristic core surface density is close to critical, which typically occurs during the early stages of core collapse. These results demonstrate that central density alone is insufficient to characterize SIDM lensing and that its evolving mass distribution must be modelled explicitly.
\end{abstract}

%%%%%%%%%%%%%%%%%%%%%%%%%%%%%%%%%%%%%%%%%%%%%%%%%%%%%%%%%%%%%%%%%%%%%%%%%%%%%%%%%%%%%%%%%%


%%%%%%%%%%%%%%%%%%%%%%%%%%%%%%%%%%%%%%%%%%%%%%%%%%%%%%%%%%%%%%%%%%%%%%%%%%%%%%%%%%%%%%%%%%

\section{Introduction}
\label{sec:intro}

Strong gravitational lensing is a purely gravitational probe of the mass distribution between a source and the observer, particularly that of the lens \citep[][]{Kneib.etal.96, Saha.Williams.97, Koopmans.etal.06}, and gives rise to phenomena including Einstein rings, giant arcs, and multiple images \citep[][]{Schneider.etal.92, Dodelson.17}.  It can also be used to constrain the particle nature of dark matter through the abundance and internal structure of low-mass haloes \citep[see][for a recent review]{Vegetti.etal.24}.

A particularly powerful window into the nature of dark matter is provided by substructure in the lenses. This can produce the following observable effects: (1) flux-ratio anomalies due to local changes in the second derivative of the lensing potential \citep[e.g.,][]{Dalal.Kochanek.02, Nierenberg.etal.17}; (2) astrometric perturbations in the surface brightness distribution of highly magnified arcs and Einstein rings induced by local changes in the first derivatives of the lensing potential \citep[e.g.,][]{Mao.Schneider.98, Metcalf.Madau.01, Vegetti.Koopmans.09, Despali.etal.22}; and (3) time-delay perturbations between multiple images of variable sources due to changes in the lensing potential itself \citep[][]{Keeton.Moustakas.09}. The frequency with which strong lenses show any of these effects constrains the abundance and masses of substructures along the line of sight, which in turn constrains the nature of dark matter. While current constraints are still weak \citep[e.g.,][]{Vegetti.etal.14, Cyr-Racine.etal.16, Hezaveh.etal.16, Gilman.etal.20, Powell.etal.23}, the statistical power is soon expected to increase manyfold with the detection of $\calO(10^5)$ strong lenses by Euclid, LSST and Roman \citep[][]{Serjeant.14, Collett.15, Ferrami.Wyithe.24, Wedig.etal.25}.

In addition to the {\it abundance} of lensing-inferred subhalos, a complementary constraint on dark matter comes from their internal {\it density profiles}. Whereas CDM predicts that dark matter haloes follow NFW profiles \citep[][]{Navarro.etal.97}, potentially modified due to baryonic processes \citep[][]{Blumenthal.etal.86, Pontzen.Governato.12}, alternative dark matter models often predict density profiles that are markedly different. For example, Fuzzy Dark Matter predicts that the centres of dark matter haloes contain solitonic cores of uniform density, in contrast to NFW cusps \citep[][]{Schive.etal.14a, Schive.etal.14b, Veltmaat.etal.18,Liao.etal.25}. Self-interacting dark matter (SIDM) also predicts cored haloes due to inward heat conduction driven by self-interactions \citep[][]{Spergel.Steinhardt.00}. Interestingly, at later times, SIDM haloes can also undergo gravothermal core collapse, causing their central densities to increase, and potentially even leading to the formation of a massive black hole \citep[][]{Balberg.etal.02, Pollack.etal.15}. This aspect of SIDM is intriguing because a number of strong lenses have been inferred to have extremely high densities that are difficult to reconcile with standard CDM. At cluster scales, the observed incidence of galaxy-galaxy strong lensing exceeds predictions from current hydrodynamical CDM simulations by about an order of magnitude \citep[][]{Meneghetti.etal.20, Meneghetti.etal.23}. Steepening the inner density profiles of the subhaloes to $\rho\propto r^{-\gamma}$ with $\gamma \sim 2.9$ largely removes this discrepancy \citep{Dutra.etal.25}. On the scale of individual galaxies, a number of studies have constrained the density profiles of perturbing dark matter subhaloes to be in tension with CDM in that they are inferred to be too centrally concentrated \citep[][]{Minor.etal.21, Enzi.etal.25, Tajalli.etal.25, Gilman.etal.26, Tekverk.etal.26}. A particularly intriguing case is the razor-thin gravitational arc detected at radio wavelengths in JVAS 1938+666 \citep[][]{McKean.etal.25}. Detailed modelling reveals a perturber with an estimated mass of $\sim 10^6 \Msun$ \citep{Powell.etal.23} with an inferred density profile reminiscent of that of a core-collapsed SIDM halo \citep{Vegetti.etal.26, Zhang.Yu.26}.

These findings have prompted a number of studies of the gravitational lensing signatures of core-collapsed SIDM haloes. \citet{Yang.Yu.21} use $N$-body simulations to demonstrate that core-collapsed SIDM subhaloes have a larger radial galaxy-galaxy strong-lensing cross section than their CDM counterparts, suggesting that SIDM could potentially alleviate the `CDM crisis' \citep[][]{Natarajan.etal.26} exposed by the work of \citet{Meneghetti.etal.20, Meneghetti.etal.23}. \citet{Hou.etal.25} develop an analytic model for gravitational lensing by SIDM haloes, concluding that an SIDM halo becomes a stronger lens as it undergoes core collapse. Using a gravothermal fluid code, they compute the density profiles of SIDM haloes as they undergo gravothermal collapse, which they then use to compute lensing characteristics (i.e., critical curves and caustics). \citet{Mace.etal.26} study the dependence of the two-point correlation function of the effective lensing convergence field on variations in SIDM core-collapse models. They model SIDM haloes according to predictions of a gravothermal fluid code and emphasize that small variations in the assumed structure of core-collapsed subhaloes can lead to significant differences in lensing power. \citet{Tekverk.etal.26} use the semi-analytical {\tt SatGen} model \citep[][]{Jiang.etal.21, Green.etal.21} together with a model for gravothermal evolution in SIDM \citep[][]{Yang.etal.24} to generate CDM and SIDM subhalo populations, which they use to model several perturbers in observed lenses. They find that an SIDM model more frequently produces structures consistent with the data.

It is often assumed that core-collapsed SIDM haloes can be described as having a steep central cusp. This has its origin in early work on core collapse \citep[][]{LyndenBell.Eggleton.80, Balberg.etal.02}, which showed that the density profile evolves from a cored profile to one with a steep cusp. In fact, in the self-similar long mean free path (LMFP) regime, it can be shown that the density profile asymptotically tends towards a power law of the form $\rho \propto r^{-\gamma}$ with $\gamma \sim 2.2$. Such steep central density cusps are also supported by numerical $N$-body simulations of core-collapsed SIDM (sub)haloes \citep[][]{Turner.etal.21, Kong.etal.26}. This is relevant from the perspective of lensing, as several studies have explicitly shown that (sub)haloes with steeper cusps often provide better fits to the data than CDM-like structures, which is typically interpreted as favouring SIDM \citep[][]{Gilman.etal.21, Dhanasingham.etal.23, Dutra.etal.25, Natarajan.etal.26, Kollmann.etal.26}.

While it is true that the initial stages of SIDM core collapse produce a density profile that resembles a steep cusp, albeit with a constant density core of ever-shrinking size, once the central core becomes optically thick and transitions to the short mean free path (SMFP) regime, the evolution starts to deviate strongly from the self-similar LMFP solution. In particular, it develops a pronounced double core structure, with an inner dense core whose mass and size are shrinking with time, superimposed on an outer core of much lower density; this structure is no longer reminiscent of a softened cusp. In addition, the core collapse itself slows down considerably during this phase, as the core now has to lose its heat via outward diffusion, rather than via direct outward transport \citep[][]{Balberg.etal.02, Outmezguine.etal.23, vdBosch.Tokayer.26}. None of the studies mentioned above has considered this phase; studies based on $N$-body simulations are unable to model core collapse beyond core densities comparable to $\sim 10^3$ times the characteristic density $\rho_\rms$ of the initial NFW profile, which for most self-interacting cross sections is still in the LMFP regime. Both \citet{Hou.etal.25} and \citet{Mace.etal.26} used their gravothermal fluid codes to probe core collapse to similar densities.
 
Here, we present a detailed study of the gravitational lensing of a background source due to an SIDM halo that is undergoing time-resolved gravothermal collapse, from its initial NFW phase, through core formation, deep into the SMFP regime when the central core density reaches values as high as $10^{17} \rho_\rms$. The gravothermal collapse is modelled using an emulator based on the results of a new gravothermal fluid code that uses adaptive radial binning in order to resolve the steep radial gradients that develop during the SMFP phase of core collapse. The emulator is used to study the overall gravitational lensing efficacy of isolated core-collapsed SIDM haloes. We show that as a halo undergoes core collapse, its lensing efficacy does not monotonically increase with time; rather, it peaks during the early stages of core collapse when the central core becomes close to critical, but subsequently rapidly decays until it drops well below that of the initial NFW profile from which it started. Given that SIDM core collapse is typically relatively fast, SIDM haloes only manifest as efficient gravitational lenses for a relatively short amount of time. Furthermore, we explicitly demonstrate that their lensing behaviour is poorly represented by that of a double-power-law density profile with a central cusp slope $\gamma > 2$.

This paper is organized as follows. Section~\ref{sec:model} describes the gravothermal fluid model that we use to compute the detailed temporal structure of SIDM haloes undergoing gravothermal core collapse. Together with a set of scaling relations, these models are used to construct a publicly available emulator that yields the full temporal evolution (including density and velocity dispersion profiles) for any NFW halo of any mass, any concentration, at any redshift, and for any (velocity-independent, elastic) collision cross section within a large but finite range. In Section~\ref{sec:lensing}, we use this emulator to model the gravitational lensing of a background source due to an SIDM halo undergoing core collapse, and we introduce a publicly available web application that can be used to reproduce all results presented in this study. The main results are presented in Section~\ref{sec:results}, where we demonstrate the transient lensing behaviour of SIDM haloes undergoing core collapse, focusing on how the results depend on halo mass, halo concentration, lensing geometry, and cross section. Section~\ref{sec:efficacy} compares the lensing efficacy of core-collapsed SIDM haloes with that of haloes with (generalized) NFW profiles, and focuses on the stage of SIDM core collapse at which lensing efficacy is maximized. Section~\ref{sec:caveats} discusses some of the limitations and shortcomings, and we end in Section~\ref{sec:concl} with a summary and discussion.

Throughout, we assume a flat $\Lambda$CDM cosmology with $\Omega_\rmm=0.3$ and $H_0 =70\kmsmpc$, unless stated otherwise.

%%%%%%%%%%%%%%%%%%%%%%%%%%%%%%%%%%%%%%%%%%%%%%%%%%%%%%%%%%%%%%%%%%%%%%%%%%%%%%%%%%%%%%%%%%%

\section{Gravothermal core collapse}
\label{sec:model}

In order to model the gravothermal evolution of an SIDM halo, we use a gravothermal fluid (\GTFns) code,  which treats the SIDM halo as a conducting fluid. Originally developed by \citet{LyndenBell.Eggleton.80} to model the collisional evolution of globular clusters, it has since been used by numerous authors to model the evolution of SIDM haloes \citep[e.g.,][]{Balberg.etal.02, Koda.Shapiro.11, Pollack.etal.15, Essig.etal.19, Nishikawa.etal.20}. 

Throughout, we assume that each halo starts out as an isotropic, spherical NFW halo\footnote{This is a reasonable assumption for an SIDM halo as long as the time-scale for violent relaxation is small compared to the collisional relaxation time.} with density profile
\begin{equation}
   \rho(r) = \frac{\rho_\rms}{(r/r_\rms) (1 + r/r_\rms)^2}\,,
\end{equation}
with $r_\rms$ and $\rho_\rms$ the characteristic radius and density, respectively. This initial density profile is truncated to ensure that the halo has a finite mass, and we define the concentration parameter $\cvir \equiv \rvir/r_\rms$. Here, $\rvir$ is the virial radius within which the average density is $\bar{\rho} = \Delta_{\rm vir} \, \rho_{\rm crit}$, with $\rho_{\rm crit}$ the critical density for closure. Throughout this paper, we use $\Delta_{\rm vir}=200$. The truncation is done using the energy-truncation method of \citet{Drakos.etal.17} as implemented in the public IC-generation code {\tt PIANISTpy} \citep{Chiang.etal.26},
%such that the total truncated mass is $M_{\rm tot} = 2 M_{\rm vir}$. Here, $\Mvir$ is $M_{NFW}(<\rvir)$ is the total mass enclosed by the untruncated NFW halo inside $\rvir$. As shown in \citet{vdBosch.Tokayer.26} the gravothermal evolution of such a truncated halo is indistinguishable from that of an untruncated halo. The 
and the initial one-dimensional velocity dispersion as a function of radius, $v(r)$, is computed by solving the Jeans equations under the assumption of isotropy.

Assuming spherical symmetry, the gravothermal evolution of this isolated SIDM halo is described by the following set of gravothermal fluid equations that are solved on a 1D radial grid: 
\begin{eqnarray}
\frac{\partial M}{\partial r} & = & 4 \pi r^2 \rho\,, \label{GTF1} \\
\frac{\partial \rho v^2}{\partial r} & = & -\frac{G M \rho}{r^2}\,,\label{GTF2} \\
\frac{L}{4 \pi r^2} & = & -\kappa \frac{\partial T}{\partial r}\,, \label{GTF3} \\
\frac{\partial L}{\partial r} & = & -4 \pi \rho r^2 v^2 \left( \frac{\partial}{\partial t} \right)_M \ln(v^3/\rho)\,.\label{GTF4} 
\end{eqnarray}
From top to bottom, these equations express mass conservation, hydrostatic equilibrium, Fourier's law of thermal conduction, and the first law of thermodynamics. $M(r,t)$ is the  mass enclosed within radius $r$ at time $t$, $L(r,t)$ is the luminosity (i.e., the amount of heat conducted across radius $r$ at time $t$), $G$ is the gravitational constant, $\kappa$ the conductivity, and $k_\rmB T = m v^2$ a measure of the temperature of the particles of mass $m$, with $k_\rmB$ the Boltzmann constant. The dark matter is assumed to be monatomic with an adiabatic index $\gamma = 5/3$. 

We solve the gravothermal fluid equations using dimensionless units. We set $M_\rms = r_\rms = 1$ and express all densities, masses, and radii in units of $\rho_\rms$, $M_\rms$ and $r_\rms$, respectively. Here, $M_\rms \equiv 4 \pi \, \rho_\rms \, r^3_\rms$. Velocities are expressed in units of $v_\rms \equiv \sqrt{G M_\rms/r_\rms}$, such that $G$ is absorbed into the velocity normalization; the collision cross section per unit mass is expressed in units of $\sigma_\rms = 4 \pi r^2_\rms/M_\rms$, and time is expressed in units of the collisional relaxation time, given by
\begin{equation}\label{tcoll}
 t_0 = (a \rho_\rms \, \sigma_\rmm \, v_\rms)^{-1}\,,
\end{equation}
which implies $t_0/t_\rms = 1/(a\hat{\sigma}_\rmm)$, where $t_\rms = r_\rms / v_\rms = \left(4\pi G\rho_\rms\right)^{-1/2}$ is the characteristic crossing time. Here, $a = 4/\sqrt{\pi} \approx 2.26$ for elastic hard-sphere-like interactions, as assumed here \citep[][]{Reif.65, Balberg.etal.02}.

Throughout, we indicate these dimensionless quantities using
\begin{equation}
\hat{r} \equiv \frac{r}{r_\rms},\qquad
\hat{\rho} \equiv \frac{\rho}{\rho_\rms},\qquad
\hat{v} \equiv \frac{v}{v_\rms},\qquad
\hat{\sigma}_\rmm \equiv \frac{\sigma_\rmm}{\sigma_\rms},\qquad
\hat{t} \equiv \frac{t}{t_0}.\qquad
\end{equation}

A useful quantity to characterize the fluid is the (local) Knudsen number, 
\begin{equation}
\Kn = {\lambda_{\rm mfp} \over H}
\end{equation} 
where $\lambda_{\rm mfp} = (\rho \sigma_\rmm)^{-1}$ is the mean free path and $H = v / \sqrt{4 \pi G \rho}$ is the gravitational scale height (or Jeans length). Where $\Kn \ll 1$ the system is in the SMFP regime and can be treated as a standard fluid, while regions with $\Kn \gg 1$ are said to be in the LMFP regime. If the entire system is in the LMFP regime, its gravothermal evolution is self-similar when expressed as a function of $t/t_0$ \citep[][see also Section~\ref{sec:selfsim}]{Balberg.etal.02, Ahn.Shapiro.05, Koda.Shapiro.11, Yang.etal.23}. 

Strictly speaking, Fourier's law of thermal conduction is only valid in the SMFP regime, where $\Kn \ll 1$. In this regime, the conductivity is given by
\begin{equation}\label{condSMFP}
\kappa_{\rm SMFP} = \frac{3}{2} {b \over a} \, n \, k_\rmB {\lambda_{\rm mfp}^2 \over t_\rmr} = \frac{3}{2} \, b \, {k_\rmB\over m} \, {v \over \sigma_\rmm}\,,
\end{equation}
\citep[][]{Balberg.etal.02, Koda.Shapiro.11}, where $n=\rho/m$ is the number density of particles, $b=25\sqrt{\pi}/32$ is the effective impact parameter, and $t_\rmr \equiv \lambda_{\rm mfp}/(av)$ is the local collision time.  In the LMFP regime, no formal expression for the conductivity exists. We follow \citet{LyndenBell.Eggleton.80} and adopt an empirical relation of the form
\begin{equation}\label{condLMFP}
\kappa_{\rm LMFP} = \frac{3}{2} \beta \, n \, k_\rmB {H^2 \over t_\rmr}
= \frac{3}{2} \,{\beta \over \pi^{3/2}} \, {k_\rmB\over m} \, {\rho v^3 \sigma_\rmm \over G}\,,
\end{equation}
with $\beta$ a free parameter of order unity. In order to interpolate between these two regimes, we follow \citet{vdBosch.Tokayer.26} and write
\begin{equation}\label{kappa_fid}
\kappa = \left[ \kappa_{\rm SMFP}^{-\alpha} + \kappa_{\rm LMFP}^{-\alpha}\right]^{-1/\alpha}
\end{equation}
Throughout, we set $(\alpha,\beta) = (0.5, 0.9)$, for which the GTF code yields results that are in excellent agreement with those obtained using detailed numerical simulations \citep[][]{vdBosch.Tokayer.26}.

Given appropriate boundary conditions, equations~(\ref{GTF1})--(\ref{GTF4}) are solved from a set of initial radial profiles. We employ a Lagrangian scheme in which each time step consists of two stages: (i) a conduction step in which equations~(\ref{GTF3}) and~(\ref{GTF4}) are solved, and (ii) a ``revirialization'' step in which hydrostatic equilibrium (equation~[\ref{GTF2}]) is restored by adjusting the radii of the Lagrangian mass shells. The conduction step is performed using an implicit backward-Euler scheme, which circumvents numerical issues in the SMFP regime. In particular, the heat-diffusion PDE is parabolic, and its spatial discretization can produce a stiff system of ODEs when gradients become large, which is especially relevant during the late stages of core collapse when the double-core structure forms. In addition, to ensure sufficient spatial resolution deep into the SMFP regime, we adopt an adaptive radial binning scheme that is described in detail in Tokayer \& van den Bosch (in prep.). In all cases discussed below, the initial radial Lagrangian grid uses $N = 260$ radial bins, logarithmically sampling the radial range from $\hat{r}_{\rm min} = 5 \times 10^{-4}$ to $\hat{r}_{\rm max} = 200$. Bins are refined whenever $\Delta \hat{r}_i/\hat{r}_i = (\hat{r}_i - \hat{r}_{i-1})/\sqrt{\hat{r}_{i-1}\hat{r}_i} > 0.1$ and merged when $\Delta \hat{r}_i/\hat{r}_i < 0.01$\footnote{The merging is required to prevent the number of radial bins from becoming too large.}. During the late stages of core collapse, the number of radial bins employed can increase to of order $1000$. 

In what follows, we use the GTF code to compute the gravothermal evolution of a single, fiducial halo for a wide range of different collision cross sections. We then describe scaling relations that apply to these results, which we use to construct an emulator that can be used for any halo that starts out with the same dimensionless density profile.

\subsection{Fiducial halo}
\label{sec:fiducial}

As our fiducial halo, we consider an initial NFW halo of mass $M_{\rm vir} = 10^{12} \Msunh = 1.429 \times 10^{12} \Msun$ at redshift $z=0$. The halo is assumed to have a concentration parameter $c_{\rm vir} = 30$ and is truncated such that its total truncated mass is $M_{\rm tot} = 2 M_{\rm vir}$. Here, $\Mvir$ is $M(<\rvir)$, the total mass enclosed by the untruncated NFW halo inside $\rvir$. The characteristic mass, radius, density and velocity are $M_\rms = 5.79 \times 10^{11} \Msun$, $r_\rms = 7.76 \kpc$, $\rho_\rms = 0.0993 \Msun \pc^{-3}$ and $v_\rms = 567.1 \kms$, while the core crossing time is $t_\rms \equiv r_\rms/v_\rms = 13.35 \Myr$. As shown in \citet{vdBosch.Tokayer.26}, the gravothermal evolution of such a truncated halo is indistinguishable from that of an untruncated halo. 
%In what follows, we refer to this halo as our fiducial halo.

We evolve the fiducial halo using the GTF code with 29 different collision cross sections that (logarithmically) sample the range from $\sigma_\rmm = 0.3 \cmg$ to $300\cmg$. Using that $\sigma_\rms = 6.225 \cmg$, these correspond to dimensionless cross sections in the range  $0.048 \leq \hat{\sigma}_\rmm \leq 48.2$. Crucially, as discussed below, the results for this halo can be rescaled to those of any other initial NFW halo, of any mass, any concentration, and at any redshift $z$, by appropriately rescaling the collision cross section and using interpolation between the results for the 29 simulations of our fiducial halo with different $\sigma_\rmm$. Throughout, we operationally define the core radius $r_\rmc$ according to $\rho(r_\rmc) = \rho_0/2$, where $\rho_0$ is the innermost tabulated density.  The core mass $M_\rmc$ is defined as the mass enclosed within $r_\rmc$, while the core density and velocity dispersion are defined as
\begin{equation}
 \rho_\rmc= {3 M_\rmc \over 4\pi r^3_\rmc}\,, \qquad
 v_\rmc= {4\pi \over M_\rmc} \int_0^{r_\rmc} \rho(r) \, v(r) \, r^2\,\rmd r\,.
 \end{equation}
We have also experimented with an alternative core definition in which the core radius is defined as the innermost radius where $\rmd\log\rho/\rmd\log r=-2$, which yields very similar values for $M_\rmc$ and $v_\rmc$. None of the results presented depend on the specific core definition used.

During the late stages of core collapse, core densities reach extremely high values and the density profile develops extremely steep gradients at the transition from the inner core to the outer core \citep[][]{Balberg.etal.02, Feng.etal.21, Outmezguine.etal.23}. As described above, we use adaptive binning in an attempt to resolve these gradients, but the implication is that the time steps have to become extremely small in order to allow for accurate revirialization. This ultimately limits the core densities that we can achieve with the GTF code, which range from $\hat{\rho}_\rmc \sim 10^{17}$ for $\hat{\sigma}_\rmm = 0.048$ to $\hat{\rho}_\rmc \sim 10^{11}$ for $\hat{\sigma}_\rmm = 48.2$. For each $\sigma_\rmm$, we register the detailed structural properties of the halo at between $300$ and $400$ snapshots taken whenever $\rho_\rmc$ changes by the same fractional amount. Since the halo spends most of its time in the core phase, while gravothermal collapse is (typically) extremely fast, this sampling is extremely non-linear in time.  

\begin{figure*}
 \centering
 \includegraphics[width=0.96\textwidth]{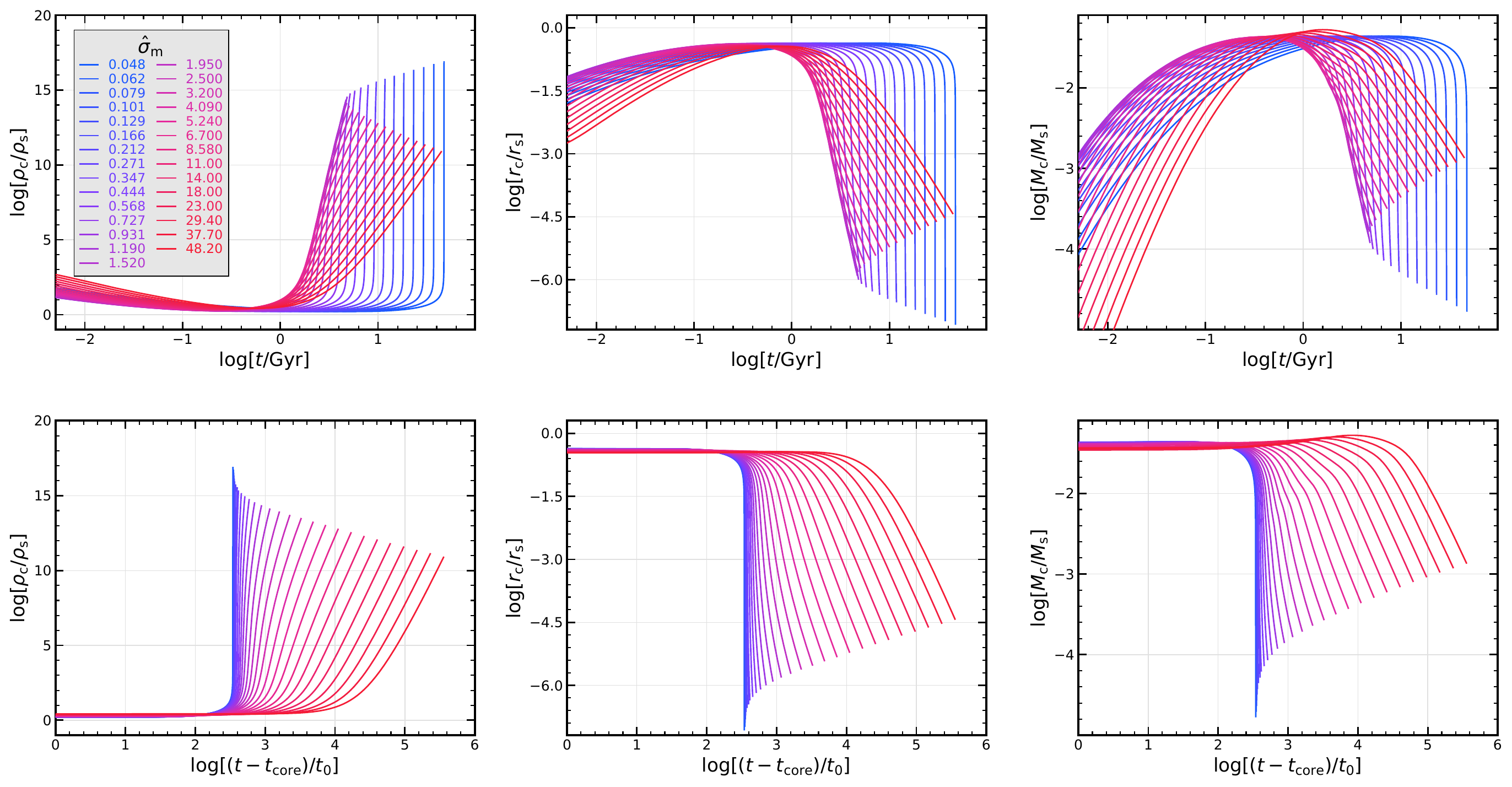}
 \caption{Results from the GTF code showing the gravothermal evolution of the fiducial SIDM halo for 29 values of the (dimensionless) cross section, as indicated. From left to right the results show the temporal evolution of central core density, core radius and core mass. Top and bottom panels show results as a function of time in Gyr and as a function of the dimensionless time variable $(t-t_{\rm core})/t_0$, respectively, where $t_{\rm core}$ is defined as the time of minimum $\rho_\rmc$, which we refer to as `maximum core'. These results are the input around which the emulator described in Section~\ref{sec:emulator} has been developed. Note the drastic decline in core size and core mass at late times, which is the key to understanding the lensing behaviour of SIDM haloes undergoing core collapse.}
\label{fig:evolution}
\end{figure*}

Fig.~\ref{fig:evolution} shows the GTF results for our fiducial halo for all 29 cross sections. From left to right, the panels show the evolution of the core density, core radius and core mass, all expressed in dimensionless form. Top panels plot the results as a function of the logarithmic time in Gyr, while the bottom panels show the results as a function of $\log[(t-t_{\rm core})/t_0]$. Here, $t_{\rm core}$ is defined as the time at which $\rho_\rmc$ reaches its minimum value, which we refer to as `maximum core'. Note the transition from evolution in the LMFP regime for small values of $\hat{\sigma}_\rmm$ to evolution in the SMFP regime when $\hat{\sigma}_\rmm$ is large. In particular, while the top panels better highlight the early stages of evolution, the bottom panels make it clear that for large cross sections the evolution deviates from the self-similar behaviour that characterizes gravothermal evolution in the LMFP regime \citep[see][for detailed discussion]{Balberg.etal.02, GadNasr.etal.24, vdBosch.Tokayer.26}. Note that the results for larger cross sections are plotted to lower central densities due to the numerical constraints mentioned above.

\subsection{Self-similarity}
\label{sec:selfsim}

The dimensionless density profile of the initial fiducial NFW halo is given by
\begin{equation}\label{NFWdim}
\hat{\rho} = {1 \over \hat{r} \, (1 + \hat{r})^2}
\end{equation}
which is only a function of the dimensionless $\hat{r}$. The same applies to the corresponding dimensionless velocity dispersion profile, $\hat{v} = \hat{v}(\hat{r})$. Hence, two haloes with the same $\hat{\sigma}_\rmm$ solve the same dimensionless evolution problem \citep[][]{Balberg.etal.02, Koda.Shapiro.11, Essig.etal.19, Yang.etal.23}. In particular, the dimensionless collision rate scales as
\begin{equation}
\hat{\Gamma} \equiv \Gamma \, t_0 = \rho \,\sigma_\rmm \, v \, t_0 = \frac{1}{a} \, \hat{\rho} \, \hat{v}\,.
\end{equation}
and the Knudsen number can be written as 
\begin{equation}
{\rm Kn} = \frac{1}{\hat{\sigma}_\rmm \, \hat{v} \, \sqrt{\hat{\rho}}} 
\end{equation}
Thus, the dimensional halo scales disappear from the collision dynamics. Gravity likewise becomes dimensionless with the above choices, and two homologous haloes with the same $\hat{\sigma}_m$ have the same Knudsen-number profile at the same dimensionless time, $t/t_0$; they therefore transition from the LMFP to the SMFP regime at the same dimensionless stage of their evolution \citep[see also][]{Outmezguine.etal.23}.

Hence, the GTF simulations for the gravothermal evolution of our fiducial halo can represent another physical halo throughout its Newtonian LMFP and SMFP evolution as long as we rescale all units consistently, preserve the complete dimensionless initial halo, and choose
\begin{equation}\label{sigma_scale}
\sigma'_\rmm = \sigma_\rmm \, \frac{\rho_\rms \, r_\rms}{\rho'_\rms \, r'_\rms}
\end{equation}
where the primes indicate the new halo\footnote{Formally, this is not true if the halo is truncated, which introduces another scale into the problem. However, as long as the truncation for the fiducial halo is sufficiently far out that its evolution is indistinguishable from that of an untruncated halo (which is the case for our fiducial halo), this is of no concern.}.

Based on equation~(\ref{sigma_scale}) and the definition of virial mass, it is straightforward to see that the dimensionless evolution of a halo $H = [\Mvir, \cvir, z, \sigma_\rmm]$, expressed as a function of the dimensionless time $t/t_0$, is identical to that of a halo $H' = [M'_{\rm vir}, c'_{\rm vir}, z', \sigma'_\rmm]$, provided that
\begin{equation}\label{mapping}
\sigma'_\rmm = \sigma_\rmm \left(\frac{M'_{\rm vir}}{M_{\rm vir}}\right)^{-1/3} \,
\left(\frac{c'_{\rm vir}}{c_{\rm vir}}\right)^{-2} \, \frac{f(c'_{\rm vir})}{f(c_{\rm vir})} \, \left(\frac{H(z')}{H(z)}\right)^{-4/3}\,.
\end{equation}
Here $f(c)=\ln(1+c)-c/(1+c)$ and we have used our adopted virial definition, according to which the virial radius encloses an average
density of 200 $\rho_{\rm crit}(z)$.

%%%Hence, the results for our library of GTF simulations with different $\sigma_\rmm$, can be used to model the gravothermal evolution of a NFW halo of arbitrary mass and arbitrary concentration at arbitrary redshift, as long as the $\sigma'_\rmm$ inferred from this mapping falls within the range covered (i.e., $0.3$ - $300\cmg$). 

\subsection{Emulator for the gravothermal evolution of NFW haloes}
\label{sec:emulator}

Using the scaling relations discussed above and the results from the GTF code for our fiducial halo with different collision cross sections, we build an emulator that uses rescaling and interpolation to compute the full gravothermal evolution of any halo $H = [\Mvir, \cvir, z, \sigma_\rmm]$ whose initial density profile is given by an (untruncated) NFW density profile. The emulator returns the temporal evolution, including radial profiles of density, velocity dispersion, and the enclosed mass at several hundred output times. More specifically, it first computes the halo’s characteristic scales $r_\rms$, $\rho_\rms$, $v_\rms$, $t_\rms$ and the dimensionless cross section $\hat{\sigma}_\rmm = \rho_\rms r_\rms \sigma_\rmm$. It then interpolates the simulated dimensionless profiles and the associated dimensionless times at the required $\hat{\sigma}_\rmm$, aligning corresponding evolutionary stages, after which it converts the resulting dimensionless radii, densities, and times into physical units using the target halo’s scales. We emphasize that the emulator assumes, and is therefore only valid for, an initial NFW profile (equation~[\ref{NFWdim}]) and is only supported within the range of dimensionless cross sections of the library of input models used (i.e., $0.048 \leq \hat{\sigma}_\rmm \leq 48.2$). 
The emulator is described in detail in Tokayer \& van den Bosch (in prep.). 
The version of the emulator used here does not extrapolate beyond the radial or evolutionary coverage of the model library, with one exception: when needed, we allow the emulator to extrapolate the density of the SIDM core to $r=0$ under the assumption that $\rmd\log\rho/\rmd\log r=0$ at radii below the innermost radial bin.

The emulator has been thoroughly tested by comparing its results to those obtained directly using the GTF code for different halo masses, concentrations, redshifts, and cross sections (see Appendix~\ref{App:emulator}). In each case, the results are found to be in excellent agreement, typically at a sub-per-cent level of precision for all emulated quantities, including the radial density and velocity dispersion profiles. The emulator is part of the publicly available lensing web application discussed below (Section~\ref{sec:lab}).

\section{Gravitational lensing}
\label{sec:lensing}

We study the gravitational lensing of a source at redshift $z_\rmS$ due to a gravothermally evolving, spherically symmetric SIDM halo (the `lens') at redshift $z_\rmL$. Throughout, we assume that the initial halo (at $t=0$) has an NFW density profile of mass $\Mvir$ and concentration $\cvir$, and we use the emulator to evolve its density profile with time, given a collision cross section $\sigma_\rmm$. For simplicity, we model the source with a circularly symmetric Plummer surface brightness profile
\begin{equation}\label{source}
 I_{\rm src}(\boldsymbol{\beta}) = I_0 \left[
 1 + \frac{|\boldsymbol{\beta} - \boldsymbol{\beta}_0|^2}{R_{1/2}^2} \right]^{-2}\,.
\end{equation}
Here, $\boldsymbol{\beta}= (\beta_x,\beta_y)$ is the true (unlensed) angular position on the sky, $\boldsymbol{\beta}_0 = (\beta_{0,x},\beta_{0,y})$ indicates the source's centroid, and $R_{1/2}$ its projected half-light radius. The origin $(0,0)$ of this coordinate system corresponds to the centroid of the lens. 

\subsection{Lensing basics}
\label{sec:basics}

Here we present a brief overview of the key aspects of gravitational lensing. For more detailed treatments, see \citet{Schneider.etal.92}, \citet{Narayan.Bartelmann.96}, and \citet{MBW10}.

Surface-brightness conservation gives the lensed image as
\begin{equation}\label{SBimage}
I_{\rm img}(\boldsymbol{\theta}) = I_{\rm src}[\boldsymbol{\beta}(\boldsymbol{\theta})]\,.
\end{equation}
Here, $\boldsymbol{\theta} = (\theta_x,\theta_y)$ is the observed angular position on the sky, which is related to $\boldsymbol{\beta}$ via the lens equation
\begin{equation}\label{lensequation}
\boldsymbol{\beta}  = \boldsymbol{\theta} - \boldsymbol{\alpha}(\boldsymbol{\theta}) - \boldsymbol{\Gamma} \, \boldsymbol{\theta} 
\end{equation}
with $\boldsymbol{\alpha}$ the reduced (or scaled) deflection angle due to the lens and $\boldsymbol{\Gamma} \, \boldsymbol{\theta}$ the contribution due to some external shear. The reduced deflection angle $\boldsymbol{\alpha}$ is related to the physical deflection angle (or bending angle) $\boldsymbol{\alpha'}$ according to
\begin{equation}\label{scaledangle}
\boldsymbol{\alpha} = \frac{D_{\rm LS}}{D_\rmS} \, \boldsymbol{\alpha'}\,.
\end{equation}
Here $D_\rmS$ and $D_{\rm LS}$ are the angular diameter distances to the source and between the lens and source, respectively. We use $D_\rmL$ to indicate the angular diameter distance to the lens. The external shear is modelled as
\begin{equation}\label{shear}
\boldsymbol{\Gamma} \, \boldsymbol{\theta} =
\begin{pmatrix}
\gamma_1 \theta_x + \gamma_2 \theta_y\\
\gamma_2 \theta_x - \gamma_1 \theta_y
\end{pmatrix}\,.
\end{equation}
Here, 
\begin{equation}\label{extshearcomp}
\gamma_1 = \gamma_{\rm ext} \cos(2\phi),\qquad \gamma_2 = \gamma_{\rm ext} \sin(2\phi)
\end{equation}
with $\gamma_{\rm ext}$ and $\phi$ the shear strength and orientation angle, and
%%%for a user-supplied shear strength $\gamma$ and shear orientation angle $\phi$, and
%
\begin{equation}
\boldsymbol{\Gamma} =
\begin{pmatrix}
\gamma_1&\gamma_2\\
\gamma_2&-\gamma_1
\end{pmatrix}
=
\gamma_{\rm ext} \,
\begin{pmatrix}
\cos2\phi&\sin2\phi\\
\sin2\phi&-\cos2\phi
\end{pmatrix}
\,.
\end{equation}

In the case of an axisymmetric lens, as assumed here, the magnitude of the scaled deflection angle due to the lens is given by
\begin{equation}\label{alphatheta}
\alpha(\theta) = \frac{2}{\theta} \, \int_0^{\theta}
\kappa(\theta') \, \theta' \, \rmd\theta'\,.
\end{equation}
Here, $\theta=|\boldsymbol{\theta}|$ and
\begin{equation}\label{convergence}
\kappa(\theta) = \frac{\Sigma(\theta)}{\Sigma_{\rm crit}}
\end{equation}
is the convergence with $\Sigma(\theta)$ the mass density of the lens projected along the line of sight and 
\begin{equation}\label{Sigcrit}
\Sigma_{\rm crit}=\frac{c^2}{4\pi G}\frac{D_\rmS}{D_\rmL D_{\rm LS}}
\end{equation}
is the critical surface density. 

%For our fiducial values of $z_\rmL$ and $z_\rmS$ we have that $\Sigma_{\rm crit} = 2079 \Msun \pc^{-2}$.

Defining the mean convergence interior to $\theta$,
\begin{equation}\label{kappa_aver}
\overline{\kappa}(<\theta) = {2 \over \theta^2} 
\int_0^{\theta} \kappa(\theta') \, \theta' \, \rmd\theta' \,,
\end{equation}
we have that
\begin{equation}\label{deflection_angle}
\alpha(\theta) = \theta \, \overline{\kappa}(<\theta)\,.
\end{equation}
Hence, the deflection at angular radius $\theta$ depends on the mean convergence inside that radius. 

\subsubsection{Critical curves and caustics}
\label{sec:caustics}

The mapping between the source plane and image plane is described by the Jacobian $\bA \equiv \partial{\boldsymbol{\beta}}/\partial{\boldsymbol{\theta}}$. The locations satisfying 
\begin{equation}\label{detA}
{\rm det}[\bA] = 0
\end{equation}
form critical curves in the image plane, which map, via the lens equation, to caustics in the source plane. Caustics divide the source plane into regions that have different numbers of lensed images \citep[see][for details]{Blandford.Narayan.86}.

In the case of a spherical lens and in the absence of external shear ($\gamma_{\rm ext}=0$), we have that ${\rm det}[\bA]=\lambda_\rmr \, \lambda_\rmt$, where 
\begin{equation}\label{tangential_eigenvalue}
\lambda_\rmt = 1 - {\alpha\over\theta} = 1 - \overline{\kappa}
\end{equation}
and 
\begin{equation}\label{radial_eigenvalue}
\lambda_\rmr = 1 - {\rmd\alpha \over \rmd\theta} = 1 - 2\kappa + \overline{\kappa}
\end{equation}
are the eigenvalues of $\bA$. 

The tangential critical curve solves $\lambda_\rmt = 0$ and is a circle of radius $\theta_\rmE$ in the lens plane, where $\theta_\rmE$ is the Einstein radius defined by $\overline{\kappa}(\theta_\rmE) = 1$. When there is no root, the lens is said to be subcritical and no Einstein radius is defined. Because $\alpha(\theta_\rmE) = \theta_\rmE$ the entire tangential critical curve maps to $\beta=0$; the tangential caustic is a point with zero enclosed area. 

In the presence of external shear, or if the lens becomes elliptical, the tangential critical curve is no longer circular, and its corresponding caustic typically takes on a four-cusped shape. In that case, we define the effective (or area-equivalent) Einstein radius as $\theta_\rmE = \sqrt{A_\rmt/\pi}$ with $A_\rmt$ the area enclosed by the tangential critical curve in the image plane.

The radial critical curves solve $2\kappa(\theta_\rmr) - \overline{\kappa}(\theta_\rmr) = 1$. If the lens is spherically symmetric and in the absence of external shear, its caustics are source-plane circles with radius $\beta_\rmr = \vert \theta_\rmr - \alpha(\theta_\rmr) \vert$. 

When the external shear $\gamma_{\rm ext}$ is non-zero, the Jacobian takes the form
\begin{equation}\label{Jaco_shear}
\bA = 
\begin{pmatrix}
1-\kappa-\gamma_{1,{\rm tot}} & -\gamma_{2,{\rm tot}}\\
-\gamma_{2,{\rm tot}} & 1-\kappa+\gamma_{1,{\rm tot}}
\end{pmatrix}
\end{equation}
with
\begin{align}
\gamma_{1,{\rm tot}} & = -h\cos2\psi+\gamma_{\rm ext}\cos2\phi\,,\nonumber \\
\gamma_{2,{\rm tot}} & = -h\sin2\psi+\gamma_{\rm ext}\sin2\phi\,.
\end{align}
Here, 
\begin{equation}
h(\theta) \equiv \bar\kappa(\theta)-\kappa(\theta)
\end{equation}
is the tangential shear generated by the lens itself, often called its internal or local shear, and $\psi$ denotes the polar angle of the image-plane position (i.e., $\theta_x = \theta \cos\psi$ and $\theta_y = \theta \sin\psi$). Hence, the Cartesian coordinates of the local shear are 
\begin{equation}
\gamma_{1,\rm halo}=-h\cos2\psi,\qquad
\gamma_{2,\rm halo}=-h\sin2\psi.
\end{equation}
which add to the components of the external shear given by equation~(\ref{extshearcomp}).

The eigenvalues of the Jacobian of equation~(\ref{Jaco_shear}) are
\begin{equation}
\lambda_\pm(\theta,\psi) = 1 -\kappa(\theta) \pm \sqrt{h^2(\theta) + \gamma_{\rm ext}^2 - 2 h(\theta) \gamma_{\rm ext} \cos[2(\psi-\phi)]}\,.
\end{equation}
which reduce to equations~(\ref{tangential_eigenvalue}) and~(\ref{radial_eigenvalue}) in the absence of external shear. Following convention, we refer to the contours $\lambda_{-}$ and $\lambda_{+}$ in the lensing plane as the tangential and radial critical curves, even though their corresponding eigenvectors are generally not exactly tangential or radial. Mapping these critical curves to the source plane via inverse ray-tracing yields the corresponding tangential and radial caustics. Numerically, we evaluate $\lambda_\pm(\theta,\psi)$ on a log-polar mesh, typically using $\sim 1000$ logarithmically spaced radii (including the origin) and $\sim 200$ polar angles. Each mesh cell is then split into two triangles, and the signs of the eigenvalues at all vertices are used to determine which edges have zero crossings. These are subsequently refined using 32 bisection iterations and joined across cells into closed curves. The zero-crossings are then mapped onto the source-plane via the lens equation to obtain the corresponding caustics. This procedure is independent of the square grid used to render the image (see below).

\subsubsection{Image construction}
\label{sec:image_construction}

In order to compute the lensed image $I_{\rm img}(\boldsymbol{\theta},t)$ at time $t$, we first calculate
\begin{equation}\label{projection}
 \Sigma(R,t) = 2 \int_R^{r_{\rm max}} \frac{\rho(r,t) \, r \,\rmd r}{\sqrt{r^2 - R^2}}.
\end{equation}
where $\rho(r,t)$ is the density profile of the SIDM halo at time $t$ and $R = D_\rmL \,\theta$ is the projected distance at the redshift of the lens. Throughout, we assume that the density is constant inside the first tabulated radius returned by the emulator and zero beyond the last tabulated radius, while using logarithmic interpolation in between.

For rendering the images, we inverse ray-trace, evaluating $\boldsymbol{\beta}(\boldsymbol{\theta})$ using equation~(\ref{lensequation}), mapping each image-plane pixel back to the source plane. We then assign the pixel its source surface brightness using equation~(\ref{SBimage}), after which we convolve the resulting image with a Gaussian PSF $\calP(\theta)$ using a Fast Fourier Transform (FFT). Throughout, we adopt a FWHM for the PSF of $\theta_{\rm FWHM}=0.05\arcsec$, which is the nominal value for the James Webb Space Telescope (JWST) at $1.5\mu m$.

The desired target pixel scale for the rendered images is taken as 
\begin{equation}
\Delta\theta_{\rm target} = \min\!\left(\frac{R_{1/2}}{5}, \frac{\theta_{\rm FWHM}}{3}\right)
\end{equation}
while the total image size $L$ is chosen to be large enough to include more than $99.9$ per cent of all the flux of the lensed image.  This implies a number of pixels $N_{\rm pix} = L / \Delta\theta_{\rm target}$, which is rounded up to a power of two. In almost all cases considered here, this yields $N_{\rm pix} = 1024$ or $2048$.
\begin{figure*}
 \centering
 \includegraphics[width=0.95\textwidth]{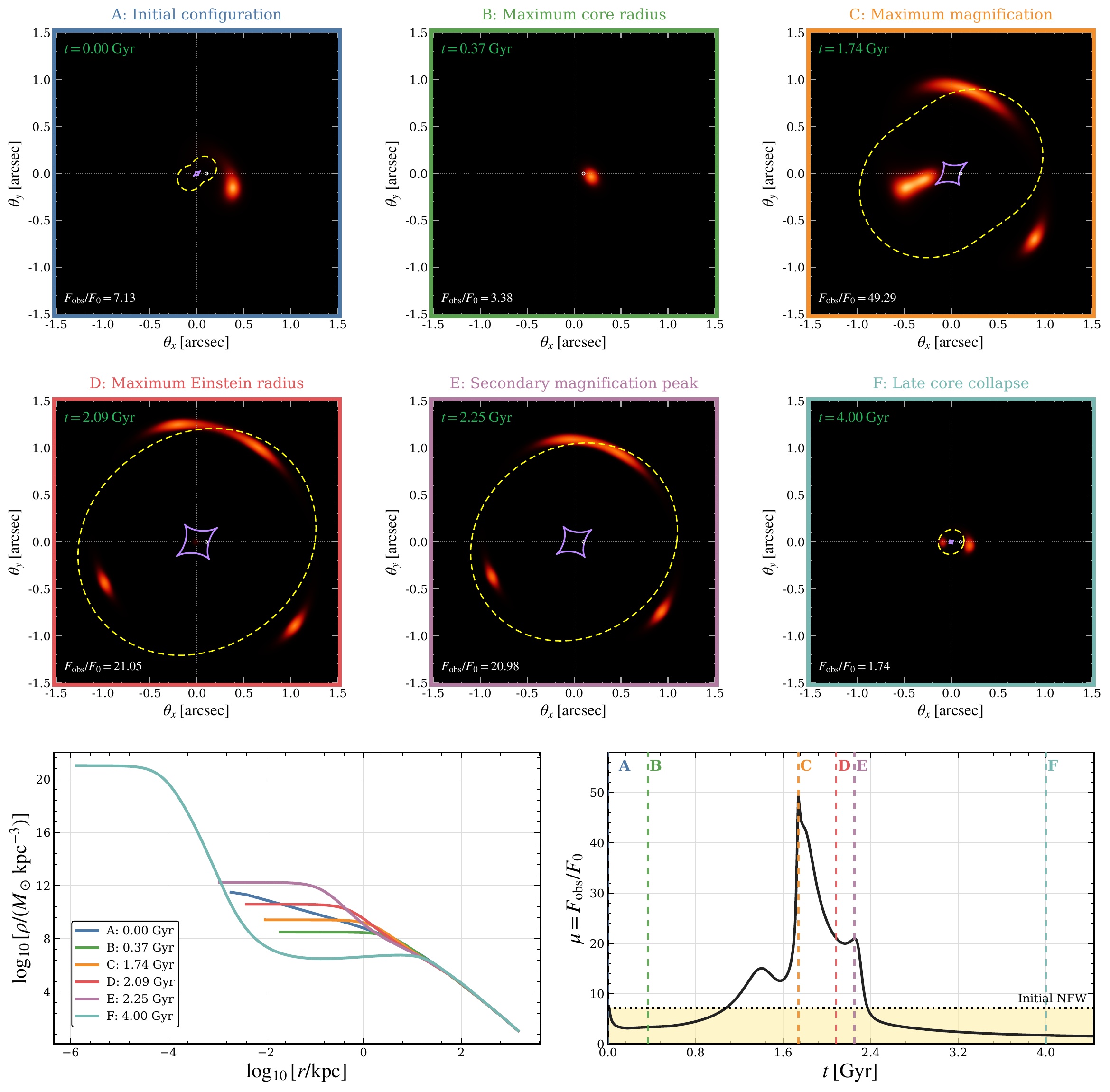}
 \caption{Six representative images of the lensed Plummer source (panels A to F) for the fiducial lensing configuration of Table~1. The panels are in chronological order and the corresponding density profiles of the SIDM lens are shown in the bottom left (colour-matched to the frame borders of panels A--F). The lower right panel shows the time evolution of the total flux magnification. Each image is labelled with the instantaneous source magnification $\mu=F_{\rm obs}/F_0$, while the small white open circle marks the position of the unlensed source. Yellow dashed and purple curves in the image panels indicate the tangential critical curve and tangential caustic, respectively. If absent, the lens is subcritical.}
\label{fig:image_sequence}
\end{figure*}

\subsection{SIDM Lensing Lab}
\label{sec:lab}

We have transformed the software developed here for the calculation of lensed images into a user-friendly app that is publicly available at \href{http://projects.frankvdbosch.com/SIDM_lensing}{http://projects.frankvdbosch.com/SIDM\_lensing}. The app allows the user to specify the mass, concentration, redshift, and collision cross section of the NFW lens, as well as the redshift and angular position of a source with respect to the projected centroid of the halo. The user can also specify the surface brightness profile of the source or can upload an image to use as a source. The app then uses the emulator to compute the lensed image of the source as a function of time as the SIDM halo undergoes its gravothermal evolution, which is displayed as a movie. The user can add an additional external shear, quantified by the shear strength $\gamma_{\rm ext}$ and orientation $\phi$, and can download the lensed images, the density, and velocity dispersion profiles of the SIDM halo as a function of time, as well as the temporal evolution of a variety of lensing parameters, including the total source magnification, the effective Einstein radius, and the areas enclosed by the radial and tangential caustics.
\renewcommand{\arraystretch}{1.05}
\begin{table}
\label{tab:fiducial}
\centering
\renewcommand{\arraystretch}{1.1}
\begin{tabular}{ll}
\hline
\multicolumn{1}{l}{Fiducial lensing configuration} \\ 
\hline
\noalign{\vskip 3pt}
Halo virial mass &   $M_{\rm vir} = 10^{12} \Msunh$ \\
Halo concentration & $c_{\rm vir} =30$ \\
Collision cross section & $\sigma_\rmm = 10 \cmg$ \\
Dimensionless cross section & $\hat{\sigma}_\rmm = 1.71$ \\
Redshift of lens  &  $z_\rmL = 0.1$\\
Redshift of source & $z_\rmS = 0.5$ \\
Half-light radius of Plummer source & $R_{1/2}=0.015\arcsec$ \\
FWHM of Gaussian PSF & $\theta_{\rm FWHM} = 0.05\arcsec$ \\
Angular source position & $\boldsymbol{\beta}_0=(0.1\arcsec, 0.0\arcsec)$ \\
External shear strength & $\gamma_{\rm ext} = 0.1$ \\
External shear orientation & $\phi = 125^{\rm o}$ \\
\hline
\end{tabular}
\caption{Parameters of the fiducial lensing configuration used throughout the text.}
\end{table}

All calculations presented below can be reproduced using the app.

\section{Results}
\label{sec:results}

As a starting point for our explorations, we consider the lensing configuration detailed in Table~1: the lens is our fiducial SIDM halo with virial mass $\Mvir=10^{12} \Msunh$ and concentration $\cvir = 30$ located at $z_\rmL = 0.1$. We assume a collision cross section of $\sigma_\rmm=10\cmg$, corresponding to $\hat{\sigma}_\rmm = 1.71$, and a circular Plummer source at $z_\rmS=0.5$ with size $R_{1/2}=0.015$ arcsec positioned at $\boldsymbol{\beta} = (\beta_{x,0},\beta_{y,0}) = (0.1\arcsec,0.0\arcsec)$, i.e., the source centre is $0.1$ arcsec offset from the projected centroid of the lensing halo. In addition, we assume the presence of external shear with strength $\gamma_{\rm ext}=0.1$ and orientation\footnote{measured anticlockwise from $+x$ towards $+y$.} $\phi=125^{\rm o}$. These numbers are merely chosen as a fiducial example, with no particular object (lens or source) in mind.

The top six panels of Fig.~\ref{fig:image_sequence} show images of the lensed Plummer source during chronologically ordered stages of the gravothermal evolution of the fiducial lens. The small white circle marks the unlensed source position, while the yellow dashed and violet curves indicate the tangential critical curve and tangential caustic, respectively. When absent, the lens is subcritical. The lower-left panel of Fig.~\ref{fig:image_sequence} shows the density profile of the SIDM halo lens at the six stages depicted in the images. The line colours match the frame colours of the corresponding images. Finally, the lower-right panel shows the temporal evolution of the total source magnification, 
\begin{equation}\label{magnification}
\mu(t) \equiv \frac{\Fobs(t)}{\Fzero}
\end{equation}
Here,
\begin{equation}\label{Flux0}
\Fzero = \int I_{\rm src}(\beta) \, \rmd^2 \beta = \pi I_0 \, R_{1/2}^2\,,
\end{equation}
is the unlensed Plummer flux integrated over the infinite source plane, and
\begin{equation}\label{Fluxobs}  
\Fobs(t) = \int I_{\rm img}(\boldsymbol{\theta},t) \, \rmd^2\boldsymbol{\theta}\,,
\end{equation}
is the total source flux integrated over the lensed image. Note that $\mu(t)$ is computed using separate adaptive integration, rather than by summing the $N_{\rm pix} \times N_{\rm pix}$ grid used to display the lensed image (see Appendix~\ref{App:magn} for details). The horizontal dashed line shows the magnification due to the initial (at $t=0$) NFW halo. Yellow shading is used to mark where the SIDM halo underperforms relative to its initial NFW halo in terms of total source magnification. The coloured vertical dashed lines mark the epochs of the six images, again colour-matched to those of their frames. 
\begin{figure*}
 \centering
 \includegraphics[width=\textwidth]{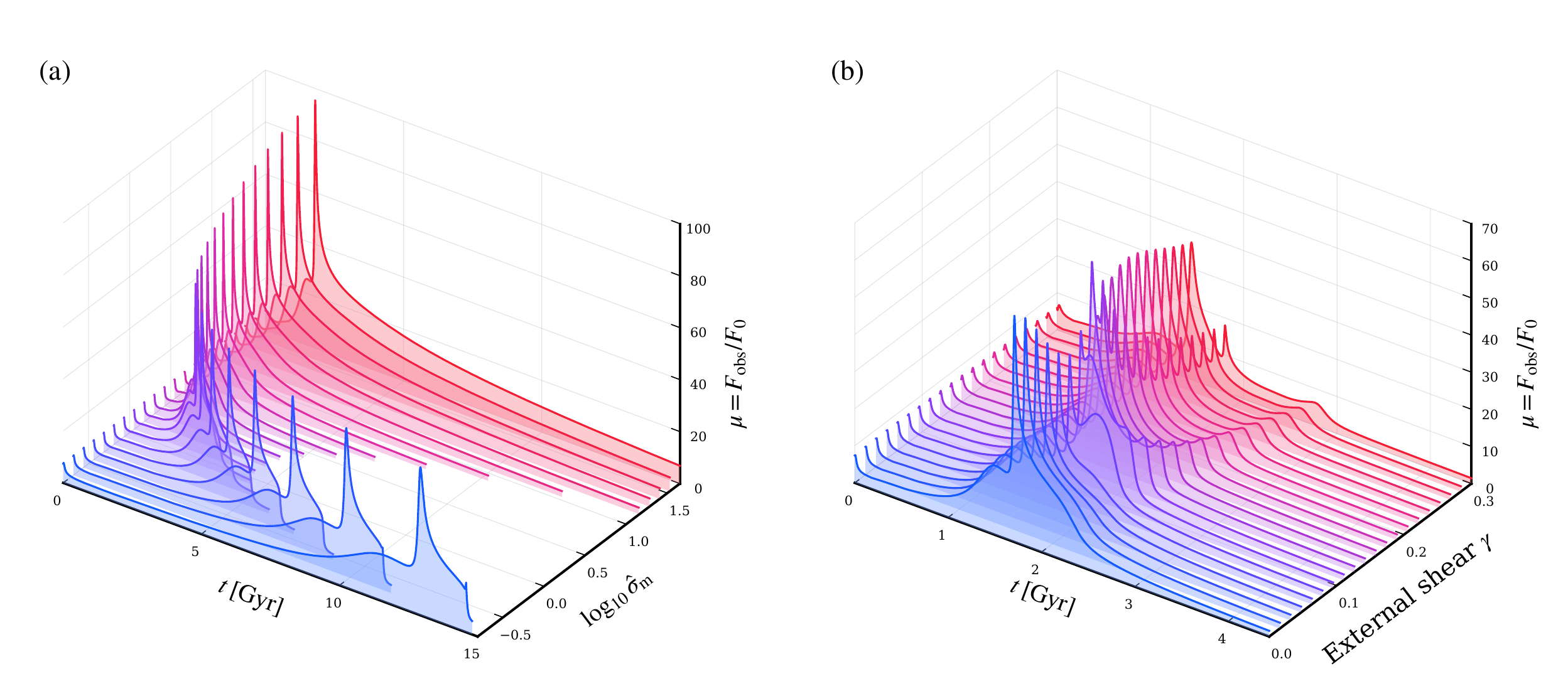}
 \caption{Panel (a) shows the temporal evolution of the total source magnification for our fiducial lensing configuration without external shear ($\gamma_{\rm ext}=0$) for 20 different values of the (dimensionless) collision cross section. Panel (b) shows the results for $\sigma_\rmm = 10\cmg$ but for different values of the external shear, which is always assumed to have the same orientation. Note the universality of the late-time decline of the magnification.}
\label{fig:magn_cross_section}
\end{figure*}

Panel A shows the initial state of the SIDM halo when it is a pure NFW halo. The Plummer source has been slightly sheared, and the total flux is magnified by a factor of $7.13$, but the source lies outside the tangential caustic, and no multiple images are produced. Panel B shows the situation at $t=0.37\Gyr$ (maximum core), when the self-interactions have created a large isothermal core in the central region of the lens, which has now become subcritical (no critical curves). The central core density has dropped to its lowest value of $\rho_\rmc = 0.32 \Msun/\pc^3$, and the magnification has decreased to $\mu = 3.38$. Panel C ($t=1.74\,\Gyr$) corresponds to the time of maximum magnification ($\mu = 49.29$), when the tangential caustic crosses over the source. The lensed image now reveals multiple images, some of which are blurred together. As is evident from the lower-right panel, core collapse has commenced but is still in its early phase, with the core density only a factor of $\sim 8.4$ higher than at maximum core. Panel D shows the situation $0.35\Gyr$ later, when the effective Einstein radius reaches its maximum value. Interestingly, although the central density has continued to increase to $39.2 \Msun/\pc^3$, the total source magnification has dropped by more than a factor of two, to $\mu = 21.05$. Panel E depicts the situation another $0.16\Gyr$ later. The central core density has now reached $1740 \Msun/\pc^3$, but both the tangential critical curve and caustic have decreased in size. In fact, at this point in time, the tangential caustic crosses the source inward, causing another peak in magnification, albeit much weaker than that at $C$. Finally, panel F shows the situation at $t=4.0 \Gyr$. Although the core density has now reached a staggering $1.02 \times 10^{12} \Msun/\pc^3$, the effective Einstein radius, tangential caustic, and source magnification have all dropped to values smaller than they were for the  initial NFW halo.

This sequence of events highlights one of the main results of this paper: {\it as an SIDM halo undergoes core collapse, its lensing efficacy does not increase monotonically.} Crucially, the mere fact that the central core density continues to increase and can reach extreme values (and perhaps ultimately even collapse to produce a black hole) does not imply that a core-collapsed SIDM halo is effective in causing strong gravitational lensing. In Section~\ref{sec:efficacy}, we quantify this using a variety of measures for the efficacy of strong gravitational lensing. However, we first take a closer look at the magnification curves resulting from the gravothermal evolution of the lens.

\subsection{Magnification curve demographics}
\label{sec:magnification}

Although the time-scale of gravothermal evolution of individual haloes is far too long for changes to be observable in a single lens, the statistics of the magnification curves are nevertheless useful in describing the population statistics of SIDM lenses. Since different dark matter haloes assemble at different times and through different merger histories, even haloes of the same mass and concentration sample different stages of core collapse and thus also different epochs of the magnification curve. Together with an assumed halo-age distribution, $\mu(t)$ determines a probability distribution for magnification: more specifically, if $P_{\rm age}(t)$ is the age distribution of the SIDM haloes under consideration, then the probability density for a magnification $\mu_{\rm obs}$ is given by
\begin{equation}
P(\mu_{\rm obs}) = \int P_{\rm age}(t) \, \delta[\mu(t) - \mu_{\rm obs}]\,\rmd t\,.
\end{equation}

One of the key insights from the fiducial lensing configuration shown in Fig.~\ref{fig:image_sequence}  is that the lensing magnification during the late stages of SIDM core collapse is low, significantly lower even than that of the initial NFW halo. Magnification generally peaks during caustic crossings. This poses the question: to what extent does the time dependence of the magnification depend on source properties (mainly size and position) versus lens properties (mass, concentration, and collision cross section)? In order to gain some insight, we examine how the magnification curve changes with several source and lens properties.

\subsubsection{Dependence on collision cross section}
\label{sec:cross_section}

The left-hand panel of Fig.~\ref{fig:magn_cross_section} shows how the time evolution of the total flux magnification depends on the dimensionless collision cross section $\hat{\sigma}_\rmm$. Here, for simplicity, we ignore any external shear ($\gamma_{\rm ext}=0.0$), but other lens and source parameters are kept identical to those of the fiducial configuration of Table~1. Several trends are evident. First and foremost, in every case $\mu(t)$ shows a sharp peak followed by a pronounced decline, highlighting the strongly transient lensing behaviour associated with SIDM core collapse, and above all, that the magnification becomes weak during the late stages of core collapse. Second, the peak occurs later for smaller values of $\hat{\sigma}_\rmm$, which is simply a manifestation of the fact that with smaller cross sections, it takes longer for core collapse to commence.

\subsubsection{Dependence on external shear}
\label{sec:shear}

The right-hand panel of Fig.~\ref{fig:magn_cross_section} shows the same flux magnification curves as in the left-hand panel, but this time for the fiducial  collision cross section of $\sigma_\rmm=10 \cmg$ and different strengths of the external shear $\gamma_{\rm ext}$. The orientation of the external shear is always kept fixed at $\phi = 125^{\rm o}$, in agreement with that used for Fig.~\ref{fig:image_sequence}. As is evident, adding external shear results in more complicated magnification curves, with multiple bumps and peaks, but the peak magnification is only weakly affected. Adding external shear creates a tangential caustic, and the source can now experience multiple caustic crossings. The richer structure evident in the magnification curves for non-zero shear is a manifestation of this more complicated caustic structure in the source plane.

\subsubsection{Dependence on source position}
\label{sec:sourcepos}

\begin{figure}
 \centering
 \includegraphics[width=0.45\textwidth]{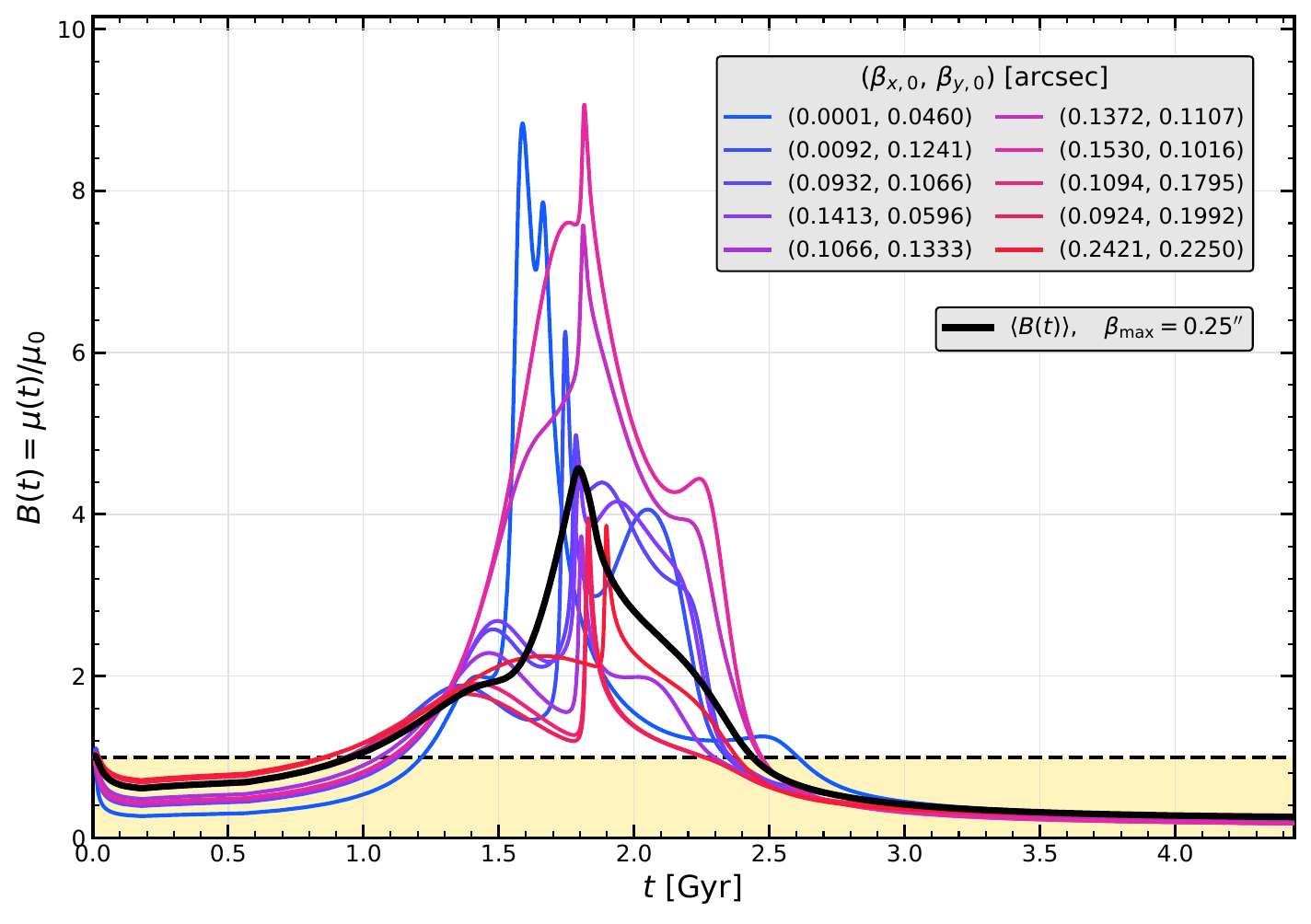}
 \caption{Boost factor evolution curves for the fiducial lensing configuration of Fig.~\ref{fig:image_sequence} for ten random source positions sampled over the intervals $\beta_{x,0} \in [0,0.25]$ and $\beta_{y,0} \in [0,0.25]$. The thick black curve shows the average magnification boost (equation~[\ref{aver_boost}]), averaged over a circular aperture of radius $\beta_{\rm max}=0.25''$. Note that the boost factor always drops below unity at both early times (formation of isothermal core) and late times (deep into core collapse).}
\label{fig:boostfactors}
\end{figure}

In order to probe the sensitivity of $\mu(t)$ to source position, we use once more the fiducial lensing configuration of Table~1, but we draw ten random source positions $\boldsymbol{\beta}_0$ from the intervals $0 \leq \beta_{x,0} \leq 0.25''$ and $0 \leq \beta_{y,0} \leq 0.25''$. Fig.~\ref{fig:boostfactors} shows the corresponding magnification boost factors $B(t)= \mu(t)/\mu_0$, where $\mu_0 = \mu(t=0)$ is the magnification due to the initial NFW halo.

The solid black line shows the average magnification boost,
\begin{equation}\label{aver_boost}
\langle B(t) \rangle \equiv {1 \over \pi \beta^2_{\rm max}} \int_{|\boldsymbol{\beta}_0|<\beta_{\rm max}} \rmd^2\boldsymbol{\beta}_0 \, {\mu(\boldsymbol{\beta}_0,t) \over\mu_0(\boldsymbol{\beta}_0)}\,.
\end{equation}
Here, $\beta_{\rm max} = 0.25''$ is the angular radius of the circular aperture in the source-plane area over which the averaging is performed\footnote{Note that this circular aperture is different from the square aperture used to draw the ten random source positions.}, $\mu(\boldsymbol{\beta}_0,t)$ is the flux magnification due to the SIDM halo at time $t$ if the  source is positioned at $\boldsymbol{\beta}_0$, and $\mu_0(\boldsymbol{\beta}_0) = \mu(\boldsymbol{\beta}_0,0)$ is the corresponding magnification due to the initial (at $t=0$) NFW halo.

Note that although there are large variations in the maximum boost and the detailed functional form of $B(t)$ near its maximum, they all fall below unity at both early times, when the SIDM halo develops a large isothermal core, and late times, when the central region develops the core-within-a-core structure with an ever-decreasing core mass and size. The same holds for the aperture-averaged boost factor $\langle B(t) \rangle$, indicating that these features are characteristic of the lens rather than the source.

Computing $\langle B(t) \rangle$ can be computationally demanding. In principle, it involves a four-dimensional integration over the image plane and the source plane, each time mapping the image plane to the source plane via the lensing equation~(\ref{lensequation}), and using sufficiently fine integration to resolve the rapidly varying magnifications near caustics. However, in the absence of external shear, and when using a spherically symmetric lens, source, and aperture, this can be simplified to a one-dimensional integral that is easily evaluated. In the case shown here, the external shear is not zero, and a more elaborate technique is needed. However, it still only took a few minutes to compute $\langle B(t) \rangle$ with accuracy at the sub-per-cent level (see Appendix~\ref{App:aperture_averaging} for details).

\subsubsection{Dependence on halo mass and concentration}
\label{sec:massconc}

\begin{figure*}
 \centering \includegraphics[width=\textwidth]{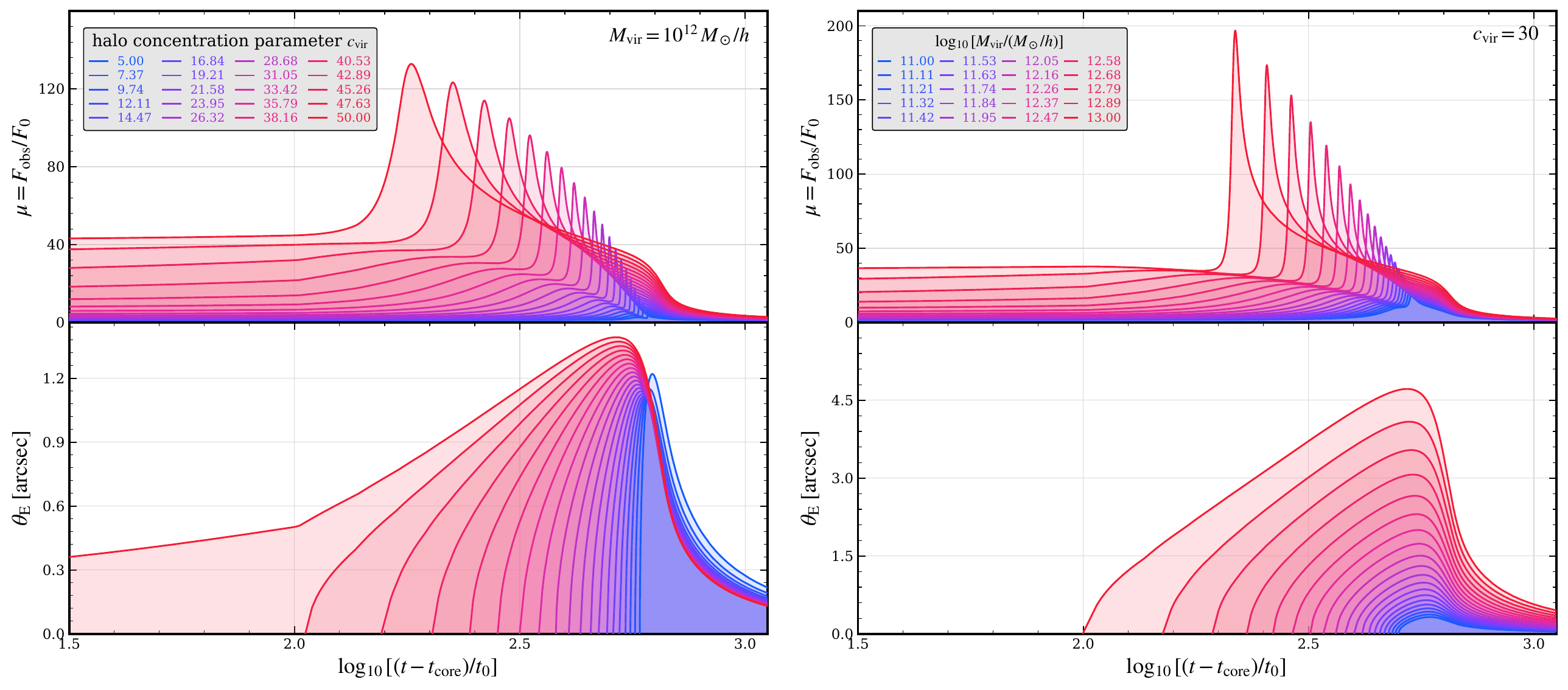}
 \caption{Dependence of the time evolution of total source magnification (top) and Einstein radius (bottom) on halo concentration (left) and halo mass (right). As expected, more massive and more concentrated haloes cause higher peak magnification and have larger Einstein radii. Note that in all cases late-stage core collapse causes negligible magnification and a rapidly declining Einstein radius.}
 \label{fig:conc_mass}
\end{figure*}

The top panels of Fig.~\ref{fig:conc_mass} show the dependence of the total source magnification on halo concentration (left) and halo mass (right). Lower panels show the corresponding time evolution of the Einstein radius, $\theta_\rmE$. Here, we keep the dimensionless cross section fixed at the fiducial value of $\hat{\sigma}_\rmm = 1.71$ but remove the external shear ($\gamma_{\rm ext}=0$). Source redshift and size are kept fixed at their fiducial values (Table~1), but we scale the source offset with the halo's scale radius $r_\rms$, such that $\beta_{x,0}/\theta_\rms \equiv \beta_{x,0} D_L/r_\rms$ remains invariant. In particular, we adopt
\begin{equation}
\beta_{x,0} = 0.1'' \, \left({M_{\rm vir} \over 10^{12} \Msunh}\right)^{1/3} \, \left({\cvir \over 30}\right)^{-1}\,.
\end{equation}

In the panels on the left, halo mass is kept fixed at the fiducial value of $M_{\rm vir}=10^{12} \Msunh$, while we vary halo concentration from 5 to 50. The corresponding physical cross section varies from $139.7\cmg$ for $\cvir=5$ to $4.3\cmg$ at $\cvir=50$. In the right panel, halo concentration is kept fixed at 30 while $\log[M_{\rm vir}/\Msunh]$ is varied from $11.0$ to $13.0$. The corresponding physical cross section scales as $\sigma_\rmm \propto \Mvir^{-1/3}$ and ranges from $21.5\cmg$ at the low-mass end to $4.6\cmg$ for $\Mvir=10^{13} \Msunh$.  

Note that more concentrated haloes reach peak magnification earlier and with higher peak magnification. They also have larger Einstein radii. This simply reflects that more concentrated haloes have larger (enclosed) surface densities, causing larger convergence and thus stronger lensing. In particular, at fixed halo mass $\Sigma(R/r_s) \propto c^2/f(c)$, which grows slightly more slowly than quadratically. At fixed halo concentration $\Sigma(R/r_s) \propto M_{\rm vir}^{1/3}$, which explains the trends apparent in the right-hand panel. 

Crucially, in all cases shown, the late-time magnification rapidly decays and the Einstein radius drops precipitously, despite the ever-increasing core density of the lens.

\subsection{Invariance of magnification curves}
\label{sec:invariance}

Varying the source and lens redshifts also impacts the magnification curves. However, their impact is purely geometrical and subject to invariance. In particular, the convergence, which depends on $\Sigma_{\rm crit}$, is invariant to the mapping $(z_\rmL,z_\rmS) \to (z'_\rmL,z'_\rmS)$ as long as the lens remains fixed\footnote{If the halo remains fixed when relocating it to another redshift $z'_\rmL$, it will no longer be labelled with the same $\Mvir$ and $\cvir$ due to the redshift dependence of $\rho_{\rm crit}$.} and 
\begin{equation}\label{Sigma_crit_scaling}
\frac{D'_\rmS}{D'_\rmL \, D'_{\rm LS}} = \frac{D_\rmS}{D_\rmL \, D_{\rm LS}}\,.
\end{equation}
where a primed $D$ indicates the angular diameter distance to one of the primed redshifts (i.e., $D'_\rmS = D_\rmA(z'_\rmS)$, etc.). This also keeps the temporal evolution of the {\it physical} Einstein radius $R_\rmE(t)$ invariant, although the angular Einstein radius $\theta_\rmE \equiv R_\rmE(t)/D_\rmL$ changes.

If we define the dimensionless coordinates 
\begin{equation}
\boldsymbol{\hat{\theta}} = \boldsymbol{\theta}/\theta_\rms
\quad\quad\quad\quad\;
\boldsymbol{\hat{\beta}} = \boldsymbol{\beta}/\theta_\rms
\end{equation}
where $\theta_\rms = r_\rms/D_L$, then, for a spherical lens and 
assuming that the dimensionless shear matrix is held fixed, the solution to the lens equation
\begin{equation}
\boldsymbol{\hat{\beta}} = \boldsymbol{\hat{\theta}} \, [1 - \overline{\kappa}(<\hat{\theta},t)] - \Gamma\hat{\theta}\,,
\end{equation}
is invariant for fixed $\Sigma_{\rm crit}$. Hence, since the magnification follows from its Jacobian, the $\mu(t)$ curves shown in Figs.~\ref{fig:image_sequence}--\ref{fig:conc_mass} are invariant under the transformation $(z_\rmL,z_\rmS) \to (z'_\rmL,z'_\rmS)$ as long as the transformation satisfies equation~(\ref{Sigma_crit_scaling}), and the source location and size change according to $\boldsymbol{\beta}_0 \to s \boldsymbol{\beta}_0$ and $R_{1/2} \to s \,R_{1/2}$, where $s = D_\rmL/D'_\rmL$. 

\section{Quantifying lensing efficacy}
\label{sec:efficacy}

The experiments in Sections~\ref{sec:cross_section}--\ref{sec:massconc} show that, while the detailed structure of the source magnification as a function of time depends strongly on halo concentration, mass, cross section, external shear, and source location, they all share some key characteristics: at early times, when the halo develops its isothermal core, the magnification is weaker than that due to the initial NFW halo. In addition, at late times, when the halo has entered the final stages of core collapse, the magnification drops precipitously and, in almost all cases explored, ends up lower than at the initial NFW stage.  This raises the question of whether, in a statistical sense, core-collapsed SIDM haloes are actually `better' gravitational lenses than their CDM counterparts, meaning `have a larger efficacy to cause significant gravitational lensing'. 

Related to this, as mentioned in Section~\ref{sec:intro}, core-collapsed SIDM haloes are often modelled in lensing studies as haloes with a generalized NFW (gNFW) density profile
\begin{equation}
\rho(r) = {\rho_\rms \over (r/r_\rms)^{\gamma} (1 + r/r_\rms)^{3-\gamma}}
\end{equation}
with a steep central cusp (typically $\gamma > 2$). Are such density profiles adequate surrogates for core-collapsed SIDM haloes when it comes to representing their strong lensing efficacy?
\begin{figure*}
 \centering
 \includegraphics[width=\textwidth]{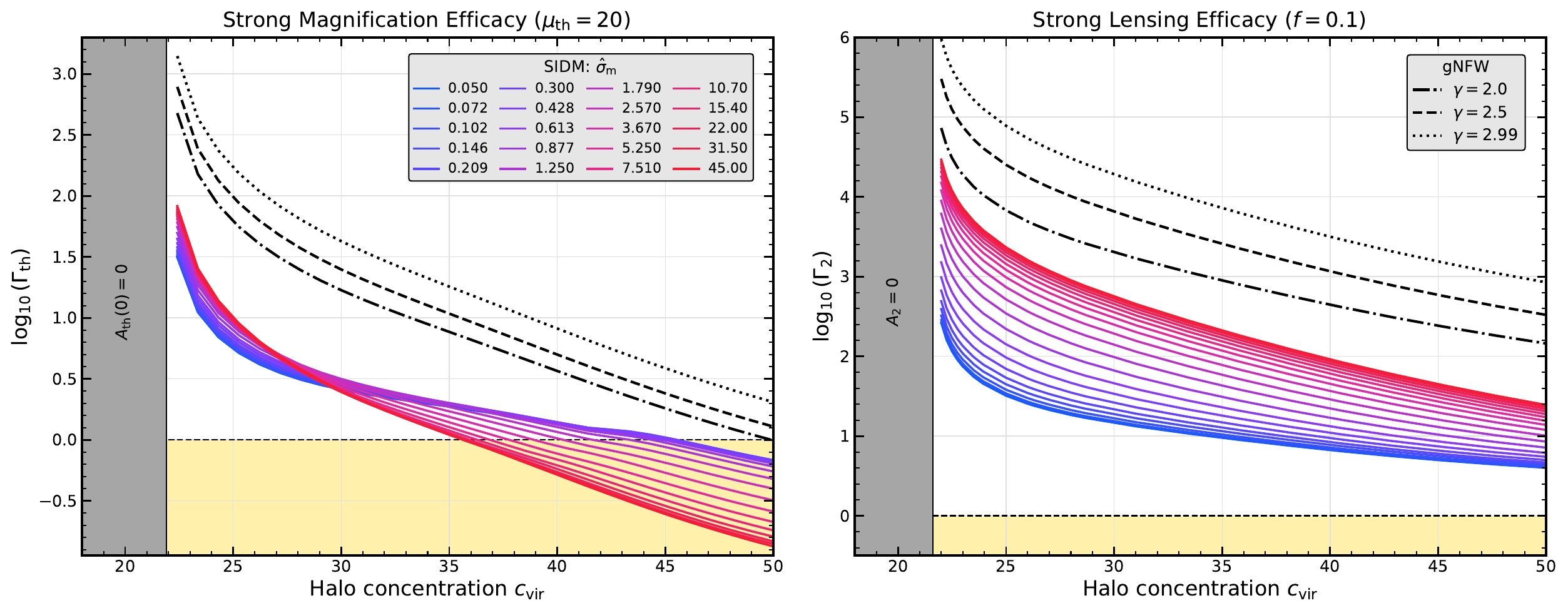}
 \caption{Gravitational lensing efficacy parameters. The coloured curves in the left and right panels show $\Gamma(\mu_{\rm th})$ for a threshold magnification $\mu_{\rm th}=20$ and $\Gamma_2$ for a minimum flux-ratio $f = \mu_2/\mu_1 =0.1$ as functions of halo concentration. Colour coding reflects the dimensionless cross section, as indicated in the left panel. The dark grey shaded areas indicate where the cross section of the reference NFW halo is zero and the efficacy parameter is undefined. The yellow-shaded region indicates where the SIDM halo has weaker lensing efficacy than the reference NFW halo. Finally, the black curves indicate the lensing efficacy of generalized NFW haloes for different values of the central cusp slope $\gamma$, as indicated in the right panel. }
\label{fig:efficacy2}
\end{figure*}

In order to address these questions, we define two measures for gravitational lensing efficacy that we use to contrast core-collapsed SIDM haloes against their NFW and gNFW counterparts (same virial mass and concentration as the $t=0$ SIDM halo). In addition, we compare these systems in terms of the structural properties of the arcs they can create. In order to simplify the computations, in what follows we set $\gamma_{\rm ext}=0$ unless specifically stated otherwise. When comparing an SIDM halo to its gNFW counterpart, we adjust the characteristic density of the gNFW halo to be 
\begin{equation}
\rho_\rms(\gamma) = {\Mvir \over 4 \pi r^3_\rms \int_0^{\cvir} x^{2-\gamma}
(1+x)^{\gamma-3} \rmd x}\,.
\end{equation}

\subsection{Strong magnification efficacy}
\label{sec:magn_threshold}

As a measure of magnification efficacy, we consider the source-plane cross section that causes a total source magnification exceeding a particular threshold value $\mu_{\rm th}$:
\begin{equation}\label{Athreshold}
A_{\rm th}(t) = \int \rmd^2\boldsymbol{\beta}_0 \, \Theta\left(\mu(\boldsymbol{\beta}_0,t) - \mu_{\rm th}\right) \,,
\end{equation}
where $\Theta(x)$ is the Heaviside step function. In what follows, we set $\mu_{\rm th}=20$ and define the magnification threshold efficacy parameter 
\begin{equation}\label{Gamma_th}
\Gamma_{\rm th} \equiv {\langle A_{\rm th} \rangle_{\Delta T} \over A_{\rm th}(0)}\,,
\end{equation}
Here, $A_{\rm th}(0)$ is the cross section for the initial (at $t=0$) NFW profile, and 
\begin{equation}\label{timeaver}
\langle A_{\rm th} \rangle_{\Delta T} = {1 \over \Delta T} \int_{T_{\rm min}}^{T_{\rm max}} A_{\rm th}(t) \rmd t
\end{equation}
is the cross section time-averaged over a period $\Delta T = T_{\rm max} - T_{\rm min}$, adopting uniform weighting in physical time. Since we are mainly interested in comparing the NFW haloes to core-collapsed SIDM haloes, we define $T_{\rm min}$ as the time of maximum core, $t_{\rm core}$, which signals the onset of core collapse. We set $T_{\rm max}$ equal to the time when the central core density reaches $10^{10} \rho_\rms$, which represents a late stage of core collapse (see Section~\ref{sec:caveats}).

We compute $\Gamma_{\rm th}$ for our fiducial lensing configuration, varying only the halo concentration and collision cross section. The results are shown in the left-hand panel of Fig.~\ref{fig:efficacy2}, which plots $\Gamma_{\rm th}$ as a function of halo concentration for 20 different values of the dimensionless cross section sampling the range $0.05 \leq \hat{\sigma}_\rmm \leq 45.0$. The dark-grey shaded region indicates the concentrations for which the initial NFW halo has zero cross section for $\mu > \mu_{\rm th}$ and thus $\Gamma_{\rm th}$ is undefined. Note that $\Gamma_{\rm th}$ typically increases sharply as the curves approach this limiting value, which reflects the drastic decline of the reference area $A_{\rm th}(0)$ rather than an increase of $\langle A_{\rm th} \rangle_{\Delta T}$. The yellow shaded region marks where $\Gamma_{\rm th}<1$, indicating that the time-averaged source-plane cross section for magnification $\mu > \mu_{\rm th}$ is smaller for the evolving SIDM halo than for its initial NFW configuration. For small concentrations, core-collapsed SIDM haloes are significantly more likely to cause a large magnification ($\mu>20$) than are their NFW counterparts. However, for large concentrations, the situation is reversed, especially at larger collision cross sections. 

In order to contrast the SIDM haloes with gNFW haloes, we also compute 
\begin{equation}\label{Gamma_th_gamma}
\Gamma_{\rm th}^{\gamma} \equiv {A_{{\rm th}, \gamma} \over A_{{\rm th}, 1}}
\end{equation}
where $A_{\rm th}^{\gamma}$ is the cross section for a gNFW halo with cusp slope $\gamma$ to cause a source magnification $\mu > \mu_{\rm th}$, and $A_{{\rm th}, 1} = A_{\rm th}(0)$ is the cross section for the corresponding (initial) NFW halo.  The results for $\gamma=2.0$, $2.5$, and $2.99$ are shown as the black lines with different line styles, as declared in the legend in the right-hand panel. This shows that in all cases, the magnification efficacy of SIDM haloes is weaker than that of gNFW haloes with $\gamma\geq 2.0$. For completeness, we mention that we have also computed $\Gamma$ for different halo masses ($10^{10.5}\Msunh \leq \Mvir \leq 10^{13.5} \Msunh$) and different threshold values ($10 \leq \mu_{\rm th} \leq 100$), always revealing very similar trends.
\begin{figure*}
 \centering
 \includegraphics[width=\textwidth]{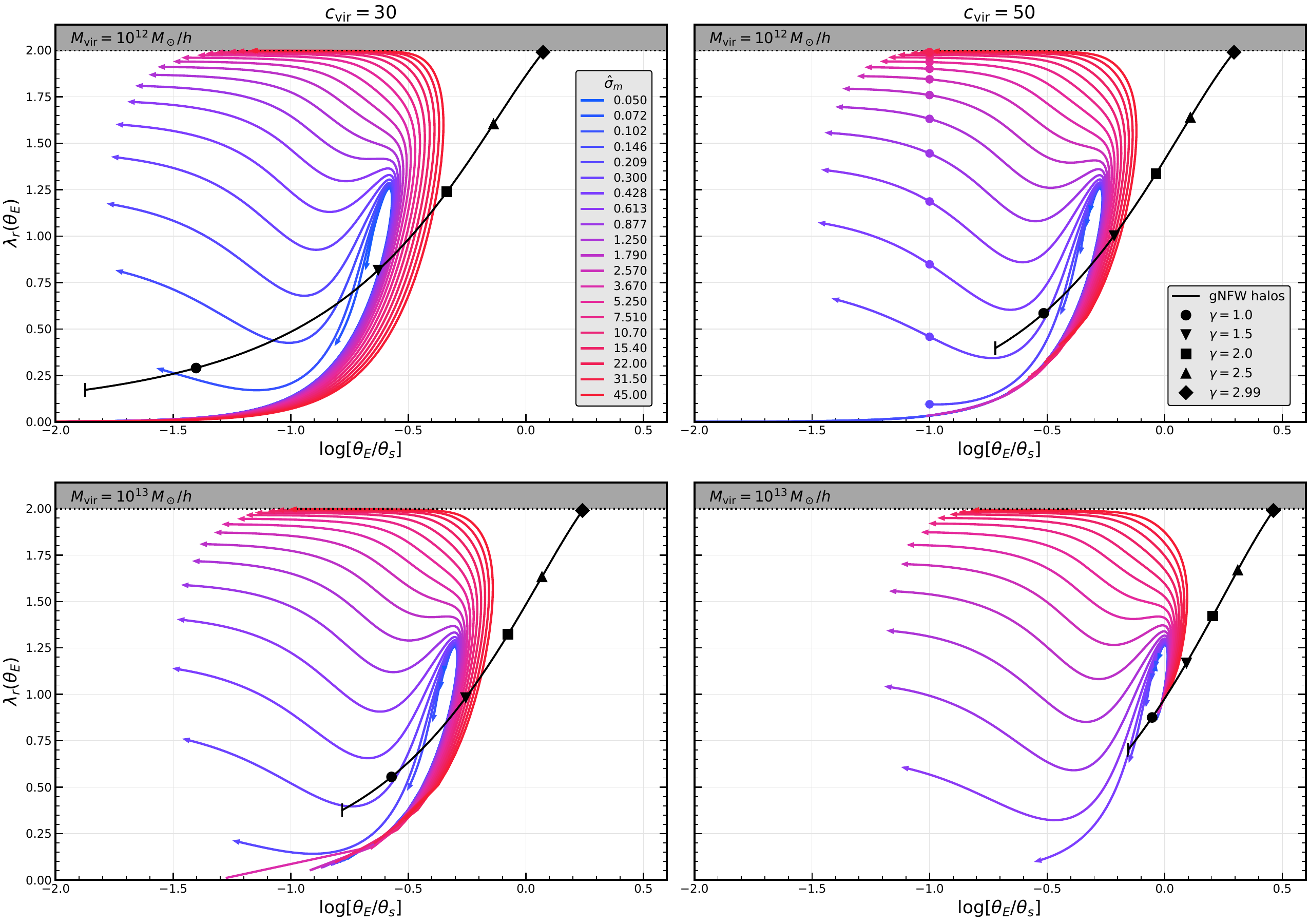}
 \caption{The radial eigenvalue at the Einstein radius, $\lambda_\rmr(\theta_\rmE)$, as a function of the Einstein radius $\theta_\rmE$ normalized by $\theta_\rms$. Coloured lines show the results for SIDM haloes for different values of the dimensionless cross section $\hat{\sigma}_\rmm$, as indicated in the top-left panel, with the arrows indicating the direction of increasing time. Top (bottom) panels show results for $M_{\rm vir} = 10^{12} \Msunh$ ($10^{13}\Msunh$), while left (right) panels correspond to a halo concentration of $\cvir = 30$ ($\cvir = 50$). The black lines show the results for the corresponding gNFW haloes with symbols marking the location for a few specific values of  $\gamma$, as indicated in the top-right panel. Finally, the grey-shaded region marks the unphysical $\lambda_\rmr>2$. The coloured solid dots in the upper-right panel mark the epochs of the density profiles shown in Fig.~\ref{fig:barSigma}, all of which have $\theta_\rmE=0.1 \theta_\rms$.}
\label{fig:Einstein}
\end{figure*}

\subsection{Strong lensing efficacy}
\label{sec:multi_eff}

The magnification lensing efficacy parameter is purely based on the source magnification. However, strong lensing is often defined as lensing that results in multiple images. In general, multiple images arise when the source is located inside either the radial or tangential caustic. For the axisymmetric lens considered here, with $\gamma_{\rm ext}=0$, the tangential caustic reduces to a point at $\boldsymbol{\beta} = 0$, while the radial caustic is a circle of radius $\beta_r$. Hence, an obvious efficacy parameter for multiple-image lensing could be the area $\pi \beta^2_r$ enclosed by the radial caustic. However, $\beta_r \to\infty$ in the limit where the lens becomes a point mass. Therefore, strictly speaking, any point-mass lens produces two images for every non-zero source offset. However, the secondary image becomes arbitrarily faint at large offset. Therefore, a meaningful efficacy parameter has to account for the observability of the secondary image \citep[][]{Oguri.Marshall.10, Lee.Kim.14, Sonnenfeld.22, Sonnenfeld.etal.23}.
\begin{figure*}
 \centering
 \includegraphics[width=0.85\textwidth]{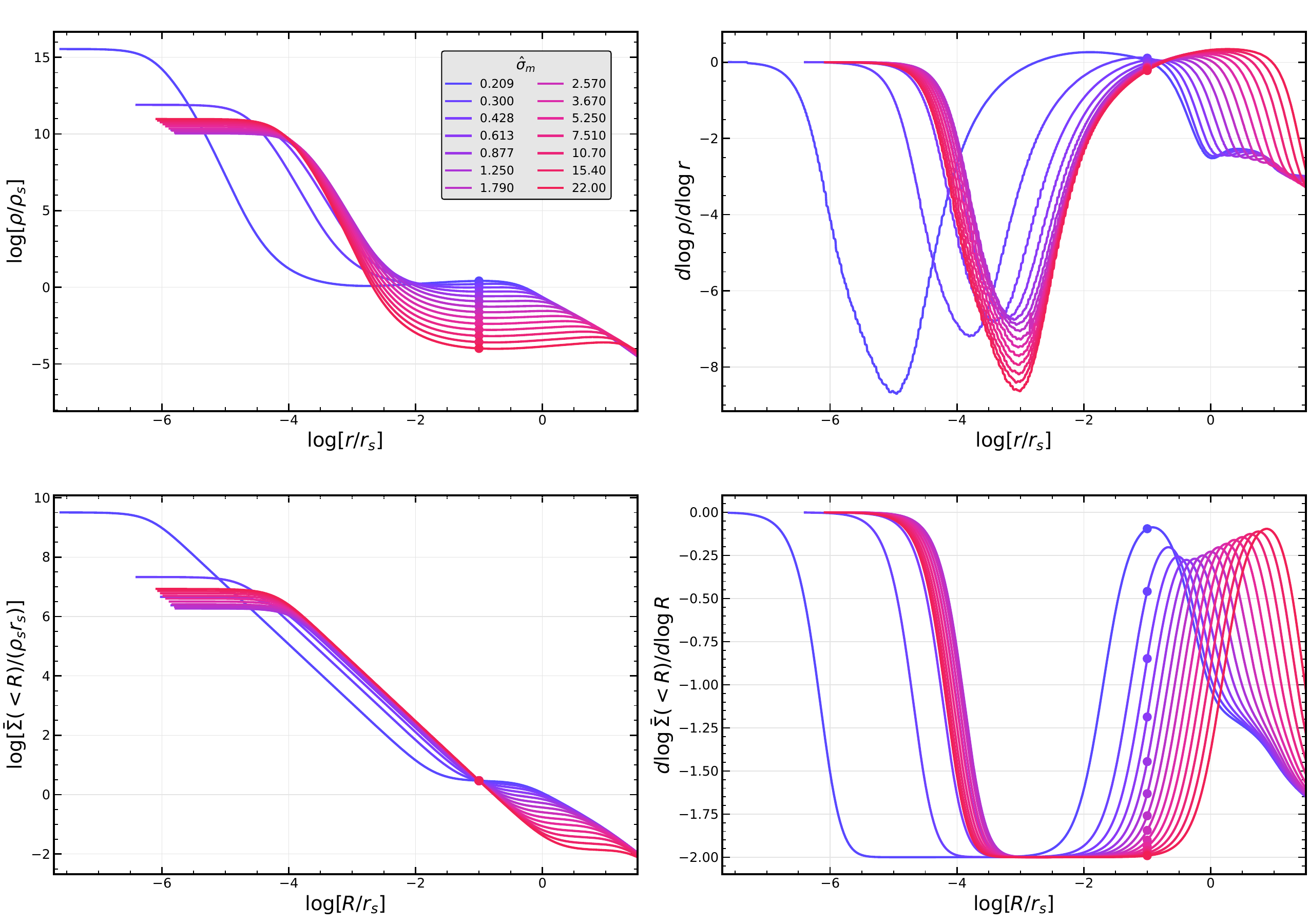}
 \caption{{\it Left panels:} The density (top) and enclosed surface density (bottom) profiles at the moment when the Einstein radius $R_\rmE=0.1 r_\rms$, as a function of 3D radius $r$ or 2D radius $R$ normalized by the scale radius $r_\rms$. All profiles are evolved from the same initial NFW halo with $M_{\rm vir}=10^{12} \Msunh$ and $\cvir = 50$, but for different values of the (dimensionless) cross section $\hat{\sigma}_\rmm$ as indicated in the top-left panel. Solid dots mark the Einstein radius at $0.1 r_\rms$. {\it Right panels:} the logarithmic gradients of the profiles shown in the left panels. Note the huge variation in $\rmd\log\bar{\Sigma}/\rmd\log R$ at $R=R_\rmE$, even though all profiles have virtually the same $\rmd\log\rho/\rmd\log r$ at their Einstein radii.}
\label{fig:barSigma}
\end{figure*}

A useful multi-image efficacy parameter is the cross section for which a second image is present with a brightness that is at least a fraction $f$ of that of the brightest image:
\begin{equation}\label{A2}
A_2(t) = \int \rmd^2\boldsymbol{\beta}_0\, \Theta\left(\mu_2(\boldsymbol{\beta}_0,t) - f \mu_1(\boldsymbol{\beta}_0,t)\right)\,,
\end{equation}
Here, $\mu_1 \geq \mu_2$ are the magnifications of images 1 and 2, ordered according to decreasing brightness at time $t$ when the source is located at position $\boldsymbol{\beta}_0$. When no second image is present, we have that $\mu_2$, and thus also the integrand, are equal to zero. In what follows, we set $f=0.1$ and define the strong lensing efficacy parameter
\begin{equation}\label{Gamma2}
\Gamma_{\rm 2} \equiv {\langle A_{2} \rangle_{\Delta T} \over A_{2}(0)}\,,
\end{equation}
where $\langle \cdot \rangle_{\Delta T}$ reflects the same time-averaging as in equation~(\ref{timeaver}), and $A_{2}(0)$ quantifies the cross section for multiple imaging for the initial NFW profile.

Note that in computing $A_2$, we use an intrinsic image-branch criterion to define the separate images (see Appendix~\ref{App:image_magn}). In reality, two images can touch at or near a critical curve or appear as a merged ring or arc, in which case an observer would have to apply some form of deblending to infer the brightness of the individual images. Such complications are not accounted for. Consequently, $\Gamma_2$ measures the efficacy of producing two sufficiently bright mathematical image branches while ignoring whether this results in two images that are resolvable as such.

The right-hand panel of Fig.~\ref{fig:efficacy2} shows $\Gamma_{2}$ as a function of halo concentration for our fiducial lens with $\Mvir = 10^{12} \Msunh$ and $\gamma_{\rm ext}=0$. Different colours correspond to different $\hat{\sigma}_\rmm$ as indicated. The dark-grey shaded region indicates the concentrations for which the initial NFW halo has zero cross section for producing a secondary image with $\mu_2 > 0.1 \mu_1$, while the yellow shaded region indicates where $\Gamma_{2}<1$. As with $\Gamma_{\rm th}$, the drastic increase in $\Gamma_{2}$ as the curves approach the critical value for the halo concentration from above reflects the vanishing of the reference area $A_{2}(0)$, rather than an increase in $\langle A_{2} \rangle_{\Delta T}$. Note that for all cross sections and all concentrations, $\Gamma_{2}$ is significantly larger than unity, by up to several orders of magnitude. Hence, a core-collapsed SIDM halo is much more likely to cause strong lensing than its NFW counterpart. 

We next consider the gNFW surrogates. The black curves in the right-hand panel of Fig.~\ref{fig:efficacy2} show $\Gamma_{2}^{\gamma} \equiv A_{2,\gamma}/A_{2,1}$ where $A_{2,\gamma}$ is the strong lensing cross section for a gNFW halo with a cusp slope $\gamma$, and $A_{2,1} = A_2(0)$ is the cross section for the initial NFW halo. This shows that a gNFW halo with $\gamma>2$ is always more likely to cause strong lensing than a core-collapsed SIDM halo. 

\subsection{The demographics of arcs and Einstein rings}
\label{sec:arcs}

Strong lensing can produce arcs or Einstein rings. These features appear at or near the Einstein angular radius $\theta_\rmE$ in the image plane. For a spherically symmetric lens, a measured Einstein radius gives a direct and tight constraint on the projected mass enclosed inside that radius. In particular, $M_{\rm 2D}(<R_\rmE) = \pi R_\rmE^2\,\Sigma_{\rm crit}$, where $R_\rmE = D_\rmL \theta_\rmE$.

Near an arc the magnification is determined by the tangential eigenvalue, $\lambda_\rmt$, while the radial eigenvalue $\lambda_\rmr$ controls the radial thickness of the arc. An important parameter, therefore, is the radial eigenvalue at the Einstein radius, which is given by
\begin{equation}
\lambda_\rmr(\theta_\rmE) = 2[1 - \kappa(\theta_\rmE)]\,,
\end{equation}
(cf. equation~[\ref{radial_eigenvalue}]), with $\kappa(\theta_\rmE)$ the convergence at $\theta=\theta_\rmE$. This, in turn, relates to the density slope of the lens near the Einstein radius. In particular, 
%if the surface density profile follows a simple power law, $\Sigma(R) \propto R^{-p}$, then $\lambda_\rmr = p$, which follows trivially from the relations given in Section~\ref{sec:basics}. More generally, 
using that
\begin{equation}
{\rmd\log\bar{\Sigma}\over\rmd\log R} = 2 \left[{\Sigma(R) \over \bar{\Sigma}(<R)} - 1 \right] = 2 \left[{\kappa(R) \over \bar{\kappa}(<R)} - 1 \right]\,, 
\end{equation}
where $\bar{\Sigma}(<R)$ is the average surface density enclosed within a projected radius $R$, it follows that
\begin{equation}
\lambda_\rmr(R_\rmE) = -\left.{\rmd\log\bar{\Sigma}\over\rmd\log R}\right|_{R_\rmE}
\end{equation}
where we have used that $\bar{\kappa}(<R_\rmE)=1$. For a point mass $\bar{\Sigma} \propto R^{-2}$ and thus $\lambda_\rmr(\theta_\rmE)=2$, which is the maximum allowed value for $\lambda_\rmr$. For a homogeneous sphere of radius $R_a$ we have that $\lambda_\rmr(\theta_\rmE) \to 0$ in the limit $R_a/R_\rmE \to \infty$. Hence, the combination of $\theta_\rmE$ and $\lambda_\rmr(\theta_\rmE)$ characterizes the demographics (size and thickness) of the arcs that result as a consequence of strong lensing, and can be used to put powerful constraints on both the enclosed mass and the slope of the density profile.
\begin{figure*}
 \centering
 \includegraphics[width=\textwidth]{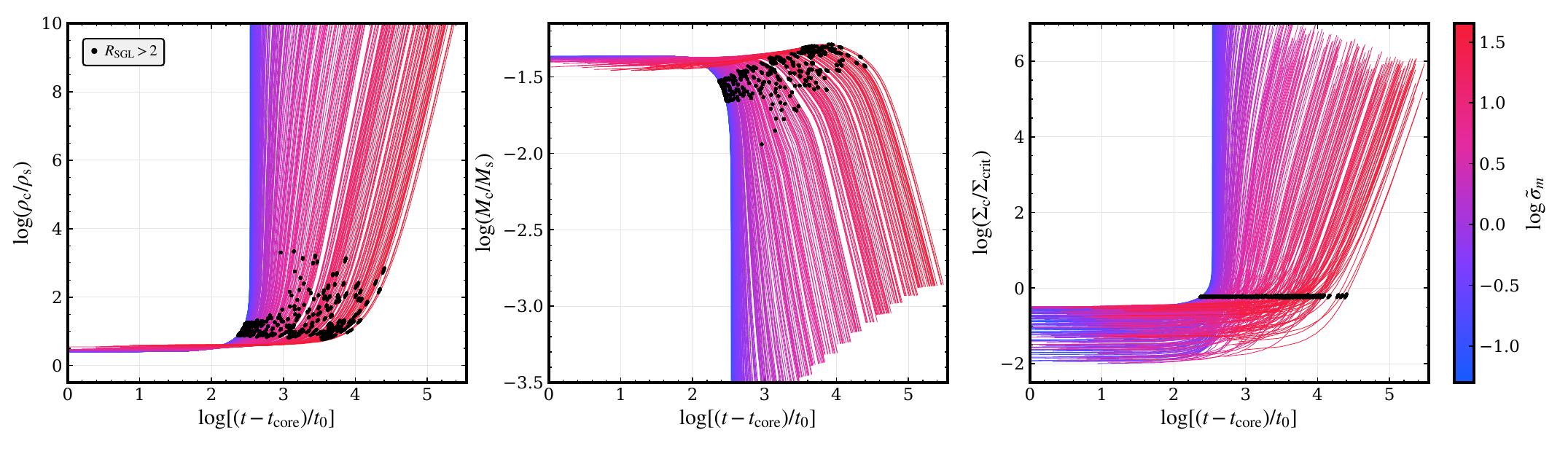}
 \caption{Evolution of the core density, core mass and core surface density for the 480 haloes of $\Mvir=10^{12}\Msunh$ with different cross sections (colour coded) and concentrations drawn randomly as described in the text. Black dots indicate the over 10,000 epochs where the cross section for strong gravitational lensing $\Asgl$ is at least two times higher than that of an NFW halo of the same mass with $\cvir=50$. Note that these epochs only pick out a very small area of parameter space, especially in terms of $\Sigma_\rmc/\Sigma_{\rm crit}$. See text for details.}
\label{fig:tracks}
\end{figure*}

Fig.~\ref{fig:Einstein} plots $\lambda_\rmr(\theta_\rmE)$ as a function of the Einstein radius $\theta_\rmE$, normalized by the angular size of the (initial) scale radius $\theta_\rms = r_\rms / D_\rmL$. Coloured curves show the temporal evolution of SIDM haloes as they undergo gravothermal core collapse. Since the haloes typically become subcritical during core formation, we start the curves in Fig.~\ref{fig:Einstein} at the first time post maximum core when the halo becomes critical. The curves end at the final times provided by the emulator, and the direction of increasing time is indicated by the arrowheads. Different colours correspond to different values of the (dimensionless) cross section $\hat{\sigma}_\rmm$, while different panels correspond to different halo masses and concentrations, as indicated. The dark-grey shaded regions mark the unphysical regime where $\lambda_\rmr>2$. 

Note that all haloes start out with rapidly increasing $\theta_\rmE$ and $\lambda_\rmr(\theta_\rmE)$. At later times, the Einstein radius reaches a maximum and starts to decrease again, typically becoming very small. This is mainly a consequence of the fact that the central core mass is decreasing with time (cf. Fig.~\ref{fig:evolution}). The late-time evolution of $\lambda_\rmr(\theta_\rmE)$ depends strongly on $\hat{\sigma}_\rmm$.  For small cross sections, $\lambda_\rmr(\theta_\rmE)$ decreases after reaching a maximum of $\sim 1.25$. For $\hat{\sigma}_\rmm \gta 0.5$ the  radial eigenvalue at the Einstein radius continues to grow in time, slowly asymptoting to a roughly constant value. For very large cross sections, $\lambda_\rmr(\theta_\rmE)$ asymptotes to the maximum allowed value of $2$, corresponding to a point mass lens. Note that these trends are qualitatively the same for all haloes, independent of their mass or concentration. In general, though, the maximum value of $\theta_\rmE/\theta_\rms$ reached increases with increasing halo mass and concentration, as already seen in Fig.~\ref{fig:conc_mass}.

It is useful to compare these results to those of gNFW haloes. The black solid lines in Fig.~\ref{fig:Einstein} indicate $\lambda_\rmr(\theta_\rmE)$ for gNFW haloes of the same mass and concentration for different values of $\gamma$. Some special cases are indicated with symbols. Note that gNFW haloes with $\gamma>2$ live in a part of $(\theta_\rmE, \lambda_\rmr(\theta_\rmE))$ parameter space that is inaccessible to SIDM haloes.  In other words, the arc demographics of gNFW haloes with $\gamma>2$ are a poor representation of those of SIDM haloes during {\it any} stage of their core collapse.

Intriguingly, the same SIDM halo can have a wide range of $\lambda_\rmr(\theta_\rmE)$ values, or a wide range of local gradients in $\bar{\Sigma}(<R)$, even when their Einstein radius, and thus $M_{\rm 2D}(<R_\rmE)$, is the same. To illustrate this, Fig.~\ref{fig:barSigma} shows the density (top) and enclosed surface density (bottom) profiles of the core-collapsed SIDM halo with $\Mvir=10^{12} \Msunh$ and $\cvir = 50$ at the moment their Einstein radii are equal to $0.1 r_\rms$, indicated with solid dots in the top-right panel of Fig.~\ref{fig:Einstein}. The left-hand panels show the profiles, while the right-hand panels show the corresponding logarithmic gradients. Note how the density at the Einstein radius is larger for smaller values of the cross section, while the logarithmic density gradients at those radii are virtually identical. For the enclosed surface density, the situation is reversed; while $\bar{\Sigma}(<R)$ is, by definition, identical, the logarithmic gradient $\rmd\log\bar{\Sigma}/\rmd\log R$ at $R = R_\rmE$ changes from almost zero for $\tilde{\sigma}_\rmm \lta 0.2$ to $-2.0$ for $\tilde{\sigma}_\rmm \gta 10$. 

At first sight, these results suggest that a given population of SIDM haloes, at different stages of their core collapse, are likely to display a much broader range of arc demographics than a similar population of CDM haloes. However, this does not account for baryons, which can drastically modify the central regions of dark matter haloes and therefore significantly impact arc demographics as well. Hence, to what extent these findings provide a pathway for testing SIDM models remains to be seen.

\subsection{The optimal SIDM lens}
\label{sec:optimal}

As discussed in Section~\ref{sec:intro}, several observational studies of gravitational lenses have inferred that the lens must be more centrally concentrated than an NFW profile. In order to put this in context with SIDM haloes, we address the following question: {\it when are SIDM haloes better lenses than a highly concentrated NFW halo of the same mass?}

As a highly concentrated NFW halo, we consider a halo of mass $\Mvir = 10^{12} \Msunh$ and a concentration parameter $\cvir=50$. We quantify its overall lensing efficacy in terms of the cross section for strong gravitational lensing $\Asgl$, defined as the area in the source plane for which the fiducial Plummer source gives rise to multiple images with $\mu_2 > 0.1\mu_1$ and {\it total} source magnification $\mu>20$, i.e., $\Asgl$ is the intersection of $A_{\rm th}$ and $A_2$. For our fiducial lensing geometry with $z_\rmL = 0.1$ and $z_\rmS = 0.5$, and adopting our fiducial Plummer source with $R_{1/2} = 0.015 \arcsec$, this reference NFW halo has $\Anfw = 0.09222\, {\rm arcsec}^2$. 

Next, we proceed as follows. We randomly draw a halo concentration from the range $5 \leq \cvir \leq 50$ and a dimensionless cross section from the logarithmic range $-1.3 \leq \log[\hat{\sigma}_{\rmm}] \leq 1.65$ (corresponding to $0.05 \leq \hat{\sigma}_\rmm \leq 45$). We then evolve an initial NFW halo of mass $\Mvir=10^{12}\Msunh$ with that concentration and cross section over time using the emulator. Adopting the fiducial source and lensing geometry, at each time step we compute the strong gravitational lensing cross section
\begin{align}\label{Asgl}
\Asgl(t) = \int \rmd^2\boldsymbol{\beta}_0 \, \Theta\left(\mu(\boldsymbol{\beta}_0,t) - 20\right) \, \Theta\left(\mu_2(\boldsymbol{\beta}_0,t) - 0.1 \mu_1(\boldsymbol{\beta}_0,t)\right)
\end{align}
and we register those epochs when $\Rsgl \equiv \Asgl(t)/\Anfw > 2$ (i.e., at which the SIDM halo has a strong lensing cross section that is twice that of its $\cvir=50$ NFW counterpart). For each halo, we also register the time $t_{\rm max}$ at which $\Asgl(t)$ reaches its maximum value, which we refer to as $\Asgl^{\rm max}$. We repeat this process until the number of haloes with $\Asgl^{\rm max} > 2 \, \Anfw$ reaches 250. Doing so requires a sample of 480 haloes and yields a total of 10,345 epochs at which $\Rsgl > 2$ (out of a total of $177,593$ epochs probed)\footnote{This implies that 5.83 per cent of all epochs probed have $\Rsgl > 2$, but we caution that, since time is not sampled uniformly, this is merely a sampling fraction and not a time-weighted fraction.}. Haloes with $\Rsgl > 2$ either have a high concentration ($\cvir \gta 35$) and/or a large cross section ($\hat{\sigma}_\rmm \gta 5$). The largest value of $\Rsgl$ found is 4.5, indicating that even the most `optimal' SIDM lens has a strong lensing cross section that is only a factor of 4.5 larger than that of a highly concentrated NFW halo of the same mass.

Fig.~\ref{fig:tracks} plots the temporal evolution of the central core density (left), core mass (middle), and the ratio $\Sigma_\rmc/\Sigma_{\rm crit}$ (right) for all 480 haloes, colour coded according to the value of the dimensionless cross section. The $10,345$ epochs at which $\Asgl(t) > 2\,\Anfw$ are marked with small black dots. Note that they only occupy a remarkably small part of parameter space; in this fiducial sample, enhancements exceeding twice the reference NFW cross section occur only during the very early stages of gravothermal core collapse. The scatter in $\Sigma_\rmc/\Sigma_{\rm crit}$ for $\Rsgl > 2$ is especially small with a median $\Sigma_\rmc/\Sigma_{\rm crit}$ of $0.604$ and a $16$-$84$ percentile range of $0.595$-$0.619$. Hence, SIDM haloes are only efficient gravitational lenses when their core surface density $\Sigma_\rmc = M_\rmc/(\pi r^2_\rmc)$ is close to 60 per cent of the critical surface density.

\begin{figure}
 \centering
 \includegraphics[width=0.5\textwidth]{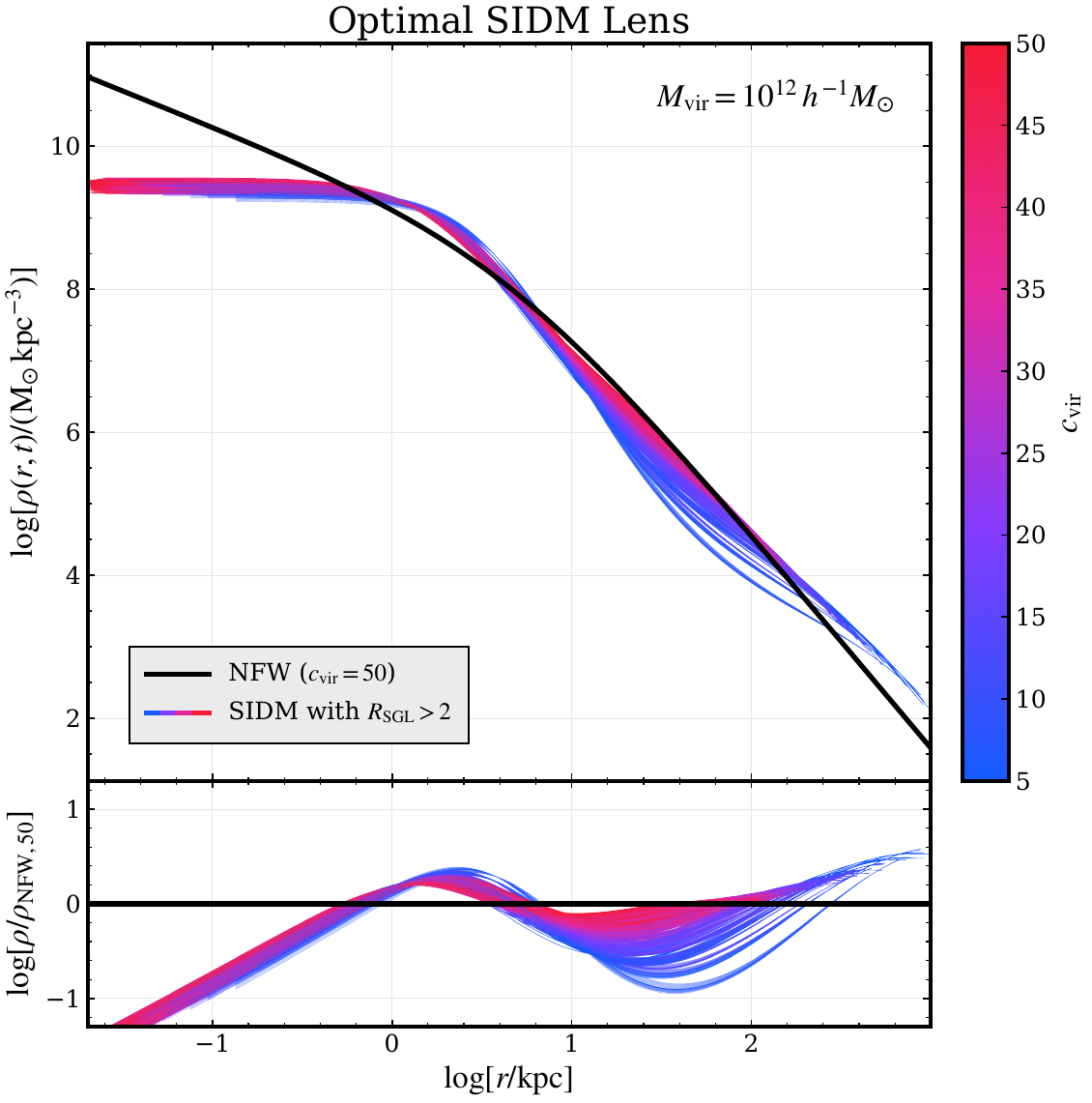}
 \caption{A collection of over 10,000 density profiles of SIDM haloes of mass $\Mvir=10^{12} \Msunh$, with a wide range of concentrations ($5 \leq \cvir\leq 50$, colour-coded) and for a wide range of cross sections ($0.05 \leq \hat{\sigma}_\rmm \leq 45$). The property all these profiles have in common is that they all have $R_{\rm SGL}>2$. Hence, these are stages of core collapse during which the halo is the optimal lens. As discussed in the text, they all have a core surface density $\Sigma_\rmc \simeq 0.6 \Sigma_{\rm crit}$.}
\label{fig:optimal}
\end{figure}

Fig.~\ref{fig:optimal} shows the density profiles of all $10,345$ epochs, in physical units, at which $\Rsgl > 2$; although these cover a wide range of concentrations and cross sections, these density profiles are remarkably similar, with a central core density close to $3 \times 10^9 \Msun/\kpc^3$. This underlines once more that there is only a very small window (in time) during which an SIDM halo undergoing core collapse is an efficient gravitational lens. For comparison, the density profile of the compact NFW halo with $\cvir=50$ is indicated as the thick black line. Note that although the latter has a higher density inside the inner kpc, the cored SIDM halo has a strong lensing cross section $\Asgl$ that is at least twice as large. 

We emphasize that reaching $\Sigma_\rmc/\Sigma_{\rm crit} \simeq 0.60$ is not a sufficient condition for being an effective gravitational lens. After all, the core surface densities of all 480 haloes probed reach a point in their evolution where this criterion is met; yet only 250 of them reach $\Asgl^{\rm max} > 2\,\Anfw$. The reason is that the core must also contain enough mass, over a sufficiently large radius, to produce a large qualifying source-plane area. Effectively, this requires that the mass enclosed by the Einstein radius $M(<R_\rmE) = \pi \Sigma_{\rm crit} R^2_\rmE \simeq (\pi/0.6) \Sigma_\rmc R^2_\rmE$ be sufficiently large, which in turn implies that the Einstein radius itself must also be sufficiently large. Indeed, $\Rsgl^{\rm max}$ is tightly correlated with $R_\rmE(t_{\rm max})$, which is well described by $\Rsgl^{\rm max} \simeq 1.85 [\theta_E(t_{\rm max})/{\rm arcsec}]^2$. Only those haloes for which the Einstein radius at $t_{\rm max}$ exceeds $1.04 \arcsec$ reach a point $\Asgl(t) > 2\,\Anfw$, which typically picks out the more concentrated haloes. It is also important to emphasize that different cross sections do not merely imply a rescaling of the evolution’s clock. As shown in \citet{GadNasr.etal.24} \citep[see also][]{vdBosch.Tokayer.26}, the structural tracks differ: at low concentration, large cross sections reach the favourable surface-density stage with substantially more core mass and a larger core radius. Hence, SIDM haloes that start out as low-concentration NFW haloes can also become efficient gravitational lenses, but only for a large enough cross section.

The analysis above focused exclusively on haloes with a virial mass of $10^{12} \Msunh$ and our fiducial lensing geometry ($z_\rmL=0.1$, $z_\rmS=0.5$) with a Plummer source. In order to test how general these results are, we perform similar tests, varying halo mass, source redshift, and lens redshift. These show that $\Sigma_\rmc/\Sigma_{\rm crit} \sim 0.6$ is a strikingly robust marker of the dominant optimal-area stage of gravothermal collapse across the entire tested mass and redshift ranges ($10^{11} \Msunh \leq \Mvir \leq 10^{13} \Msunh$, $0.1 \leq z_\rmL < 0.5$, $0.5 \leq z_\rmS \leq 2.0$). However, these tests also reveal that this stage is not unique: under certain conditions, an SIDM halo can give rise to $\Rsgl > 2$ at a much later stage of core collapse, when $\Sigma_\rmc/\Sigma_{\rm crit}$ is of the order of $10^6$--$10^7$. Typically, this requires haloes for which the initial characteristic surface density $\rho_\rms r_\rms$ is large, which requires more massive, more concentrated haloes. 

\section{Limitations and shortcomings}
\label{sec:caveats}

The previous section presents a detailed assessment of the strong lensing efficacy of SIDM haloes undergoing gravothermal core collapse. However, the results are subject to a number of limitations and shortcomings: 

\begin{itemize}[leftmargin=0.27truecm, labelwidth=0.2truecm]

\item {\it Spherical symmetry:} We have focused exclusively on spherically symmetric haloes, albeit it occasionally with external shear. Ellipticity, triaxiality, substructure, and line-of-sight structure, all of which are present in more realistic settings, change the caustic topology and magnification statistics. The perfect axial symmetry adopted here is intentional to facilitate a transparent comparison of the lensing efficacies of SIDM haloes and their gNFW counterparts. 

\item {\it Haloes in isolation:} In the case of non-isolated haloes, two effects play an important role: tidal stripping can accelerate the core collapse of an SIDM halo \citep[][]{Nishikawa.etal.20, Fischer.etal.25}, while motion through the host halo can give rise to evaporation \citep[][]{Kummer.etal.18, Zeng.etal.22}. Additionally, the presence of a host halo results in external shear, which impacts the lensing, as demonstrated here by including a simple external shear model \citep[see also][]{Hou.etal.25}. 

\item {\it Baryons:} No baryonic effects have been taken into account. Baryons impact lensing through the mass they contribute, but also through the way they modify the density profile of the halo. Baryonic processes can facilitate core formation \citep[][]{Pontzen.Governato.12}, but can also accelerate core collapse \citep[][]{Elbert.etal.18, Zhong.etal.23, vdBosch.Dattathri.26}. Hence, baryons can have an appreciable impact on all the lensing statistics explored here, something that is left for future study.

\item {\it Velocity-independent cross sections:} The gravothermal evolution of SIDM haloes explored here is based on the assumption that the collision cross section has no velocity dependence. However, many particle-physics SIDM models, especially light-mediator/Yukawa-like models, naturally predict anisotropic, velocity-dependent scattering \citep[][]{Tulin.Yu.18, Yang.Yu.21, Girmohanta.Shrock.22}. Typically, these models predict that $\sigma_\rmm \propto v^{-4}$ at large velocities, which can have an appreciable impact on the gravothermal evolution \citep[][]{GadNasr.etal.24}. On the other hand, \citet{Outmezguine.etal.23} demonstrated an approximate universality that allows velocity-dependent SIDM models to be mapped onto velocity-independent ones. Hence, since our analysis has covered a large range of (velocity-independent) cross sections, we conjecture that qualitatively our results remain valid for velocity-dependent cross sections.

\item {\it Fixed geometry and source:} Throughout, we have mainly restricted ourselves to one 'fiducial' lensing geometry with $z_\rmL =0.1$ and $z_\rmS=0.5$ and a circular Plummer source with a half-light radius of only $0.015\arcsec$. The extent of the source impacts how the caustics cross over the source, which manifests as changes in the time evolution of the magnification, while changes in $z_\rmL$ and $z_\rmS$ impact the critical surface density $\Sigma_{\rm crit}$, which, in turn, affects the convergence and caustic structure. We expect the qualitative non-monotonic evolution, which is lens-specific rather than geometry-specific, to persist, although its amplitude, timing, and associated cross sections depend on source size and geometry. This is supported by the fact that a mapping $(z_\rmL,z_\rmS) \to (z'_\rmL, z'_\rmS)$ exists that leaves the magnification curves invariant (see Section~\ref{sec:invariance}).

\item {\it Observability:} The efficacy parameters used are based on the total flux, either of the entire image or of the individual image branches, without accounting for observable limitations related to image separation, surface brightness limits, signal-to-noise, dynamic range, or image deblending.  Population observability requires an explicit survey selection function, which is left for future work.

\item {\it Finite time integration:} When computing the time-averaged lensing efficacy parameters, we do so by integrating up to a maximum time $T_{\rm max}$ which is taken to be the time when the central core density reaches $10^{10} \rho_\rms$. This choice was made because the emulator can evolve each halo up to that density, independent of cross section. An alternative would be to set $T_{\rm max}$ to the maximum time provided by the emulator, in which case haloes with smaller cross sections can be evolved to higher core densities\footnote{This is a consequence of numerical restrictions originating from the GTF calculations (see Fig.~\ref{fig:evolution} and discussion in Section~\ref{sec:fiducial}).}. We have verified that doing so yields results that are almost indistinguishable from those in Fig.~\ref{fig:efficacy2}. Ideally, we should set $T_{\rm max}$ either to infinity or, more physically, to the Hubble time at the redshift of the lens. In general, extending $T_{\rm max}$ to later times will reduce the efficacy parameters $\Gamma_{\rm th}$ and $\Gamma_2$ because, at late times, the SIDM haloes become highly inefficient lenses.

\end{itemize}

All these limitations are likely to have a quantitative impact on the results presented in Section~\ref{sec:results}. However, we conjecture that the key qualitative result --- that SIDM haloes are only effective as gravitational lenses for a relatively short period during the early stages of core collapse --- is robust, given that it is mainly a manifestation of the nature of core collapse.

\section{Conclusions}
\label{sec:concl}

In recent years, several gravitational lensing studies have revealed the existence of lenses with highly concentrated density profiles that are difficult to reconcile with the CDM paradigm. It has been argued instead that these lenses are indicative of SIDM haloes that have undergone gravothermal core collapse. This conclusion is supported by a number of recent studies arguing that as an SIDM halo undergoes core collapse, it becomes a stronger, more potent gravitational lens \citep[][]{Yang.Yu.21, Hou.etal.25, Natarajan.etal.26, Mace.etal.26, Tekverk.etal.26}. However, none of these studies has followed core collapse to core densities significantly higher than $10^3 \rho_\rms$. Core collapse is a runaway process, and the core rapidly reaches much higher densities. In fact, at some point, it will enter the SMFP regime and become optically thick, which causes a significant deviation from the self-similar evolution characteristic of the LMFP regime. In particular, the density profile develops a core-within-a-core structure, with a dense core of ever-decreasing size and mass superimposed on a much larger core of much lower density (Fig.~\ref{fig:image_sequence}).

Using an emulator that accurately models the gravothermal collapse of SIDM haloes well into the SMFP regime, we have studied the gravitational lensing of simple background sources due to SIDM haloes undergoing core collapse. The total finite-source flux magnification of a background source first decreases as the halo forms its isothermal core. It then rapidly increases as core collapse commences, after which it declines precipitously. These results are robust to changes in halo mass, concentration, and cross section, as well as to lens geometry and the presence of external shear (Fig.~\ref{fig:magn_cross_section}--\ref{fig:conc_mass}). This transient behaviour has its origin in the structural evolution of the lens. During early collapse, increasing density and sufficient mass over a source-sized region move a critical curve or caustic through the extended source, producing sharp peaks in magnification.  Later, the high-density component contracts to a much smaller radius and contains progressively less mass, while the outer halo changes more slowly.  The core can then continue to collapse without maintaining its maximum effect on rays that sample scales set by the Einstein region, source offset, and finite source size.

Because of the strong time dependence, comparing the lensing efficacy of an SIDM halo undergoing core collapse to that of a (generalized) NFW profile requires time-averaging, treating the time dependence as a probability function for observing the lens in a particular stage of its core collapse. We constructed two lensing efficacy parameters: one that measures the time-averaged cross-sectional area in the source plane for magnification above some threshold value ($A_{\rm th}$), and the other that measures the time-averaged cross-sectional area in the source plane for creating multiple images, with the flux of the second brightest image being at least 10 per cent of that of the brightest image ($A_2$). While a time-averaged SIDM lens typically, but not always, has a larger efficacy than its NFW counterpart, it always underperforms compared to a gNFW halo with a cusp slope of $\gamma \geq 2$ (Fig.~\ref{fig:efficacy2}). Hence, we conclude that halo models with a steep central cusp are poor proxies for core-collapsed SIDM haloes when it comes to modelling gravitational lenses.

As an SIDM halo undergoes core collapse, the Einstein radius increases rapidly after reaching maximum core. Then, shortly after the system achieves maximum magnification, the Einstein radius rapidly declines. Interestingly, during this decline, the radial eigenvalue at the Einstein radius, $\lambda_\rmr(\theta_\rmE)$, which reflects the logarithmic slope of the enclosed surface density at the Einstein radius and controls the radial thickness of the lensing arc, becomes extremely sensitive to the value of the SIDM collision cross section. For large cross sections, it rapidly asymptotes to the maximum value of 2, indicative of a point-mass lens, while for sufficiently small cross sections, it can drop close to zero, indicative of a homogeneous core (Fig.~\ref{fig:Einstein}). This reflects that the cross section determines during what stage of core collapse the LMFP-to-SMFP transition occurs, which breaks the self-similarity. 

Finally, since SIDM haloes are often invoked when even a highly concentrated NFW halo can no longer fit the data, we defined the cross section for strong gravitational lensing, $\Asgl$, as the intersection of $A_{\rm th}$ and $A_2$. We find that an SIDM halo undergoing core collapse only has an $\Asgl$ that is significantly larger than that of a highly concentrated NFW halo of the same mass for a surprisingly narrow period in time during the early stages of core collapse (Fig.~\ref{fig:tracks}). This stage of core collapse is characterized by a very specific density profile (see Fig.~\ref{fig:optimal}) during which the core is relatively large and has a surface density that is close to critical ($\Sigma_\rmc \simeq 0.6\Sigma_{\rm crit}$). At later stages, $\Sigma_\rmc$ increases, but both core size and core mass decrease, which drastically reduces $\Asgl$ (see Appendix~\ref{App:cores}).

Based on these tests, we conclude that SIDM haloes are not as efficient as strong gravitational lenses as often argued. While it is true that their lensing efficacy increases during the early stages of core collapse, this only lasts for a relatively short amount of time; once the core has become supercritical, the lensing efficacy of a core-collapsed SIDM halo rapidly declines and often drops below that of the original NFW halo from which it evolved. Systems with steep centrally cusped density profiles are extremely poor proxies for core-collapsed SIDM haloes; rather, they should be modelled using density profiles supported by more detailed calculations. The publicly available emulator developed as part of this project can be used to compute the full time evolution of an SIDM halo undergoing core collapse and will hopefully prove useful for such modelling. 

\section*{Data availability}

Most data presented here have been obtained using the \href{http://projects.frankvdbosch.com/SIDM_lensing}{SIDM lensing app} publicly available at http://projects.frankvdbosch.com/SIDM\_lensing and specifically constructed for this project. The app can be used to compute and download the gravothermal core collapse of SIDM haloes and to perform the majority of the lensing calculations presented here.

\section*{Acknowledgements}

We are grateful to Barry Chiang, Shashank Dattathri, Isaque Dutra, Dušan Jovanović, Sebastian Monzon, Priya Natarajan, and Simona Vegetti for useful discussions. OpenAI Codex (GPT-6 Sol and GPT-6 Astra) has been used to build the emulator and lensing app, to calculate lensing efficacy parameters, synthesize the literature, and produce all figures. The first author reviewed and verified the calculations, code outputs, references, figures, and scientific conclusions, and takes full responsibility for the work. The actual paper and the GTF code that served as input for the emulator were written by FvdB and YT without any AI-based input. This work has been supported by the US National Science Foundation through grant AST-2407063.

%\FloatBarrier

%%%%%%%%%%%%%%%%%%% REFERENCES %%%%%%%%%%%%%%%%%%

\bibliographystyle{mnras}
\bibliography{references_vdb} 

%%%%%%%%%%%%%%%%% APPENDICES %%%%%%%%%%%%%%%%%%%%%

\appendix
\numberwithin{figure}{section}
\numberwithin{table}{section}
\numberwithin{equation}{section}

\section{Testing the emulator}
\label{App:emulator}

The results in the main text rely heavily on an emulator that, given the virial mass, concentration, redshift, and collision cross section, computes the gravothermal evolution of an NFW halo, including full radial profiles for density, enclosed mass, and 1D velocity dispersion. As described in Section~\ref{sec:emulator}, the emulator is constructed using scaling relations and interpolation between the results from the GTF code for one initial halo at 29 different values of the cross section. 
\begin{figure*}
 \centering
 \includegraphics[width=0.9\textwidth]{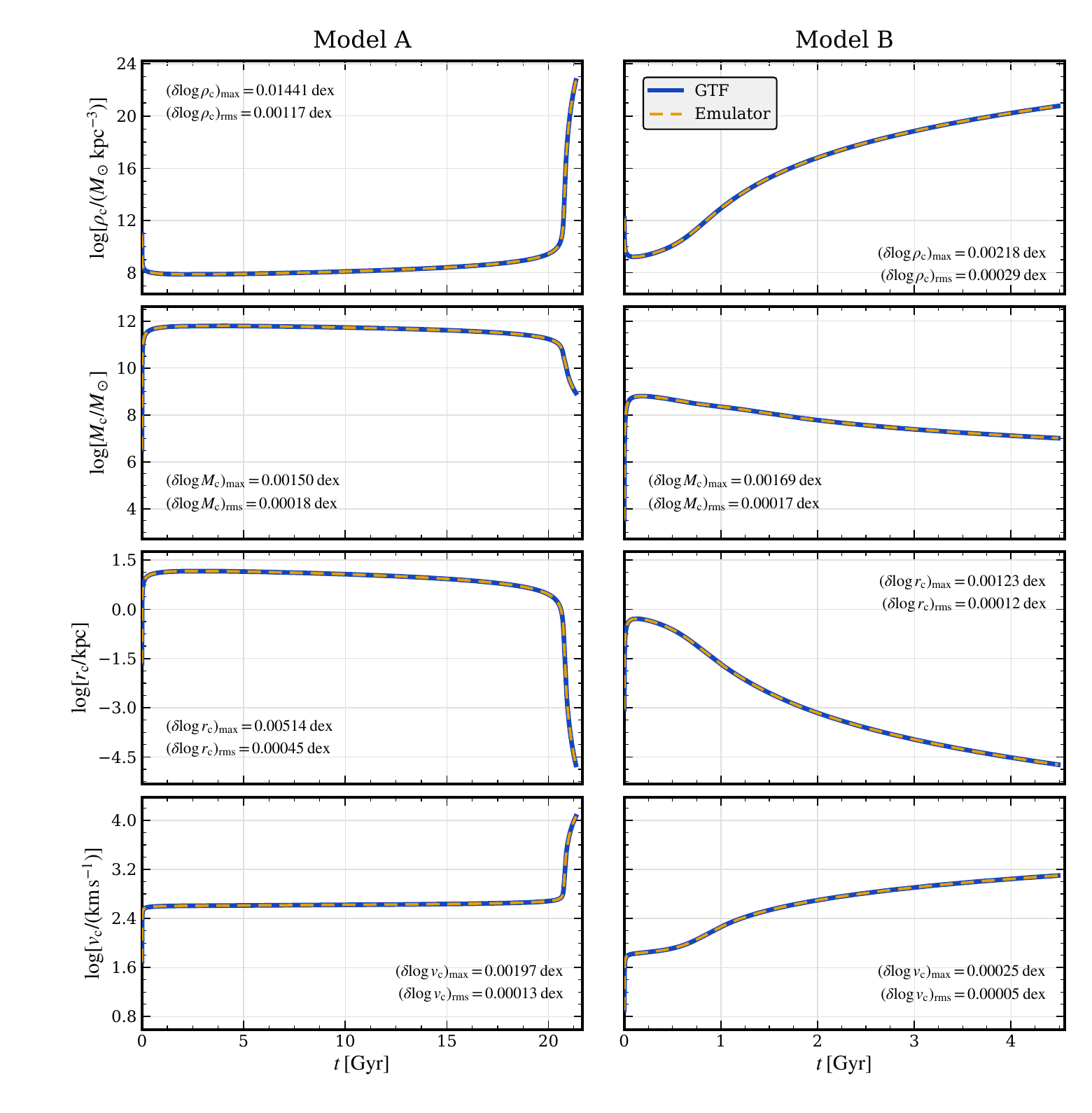}
 \caption{Time evolution of, from top to bottom, core density, core mass, core radius and core 1D velocity dispersion, for the two models A and B used to validate the emulator (see Table~A.1). The thick blue line shows the results obtained with the GTF code, while the orange dashed line shows the predictions from the emulator. The agreement is excellent, as is also evident from the maximum and rms logarithmic errors indicated in each panel.}
\label{fig:emulator_test}
\end{figure*}

Here we test the performance of the emulator using two SIDM haloes, A and B, whose properties are listed in Table~A.1. They differ substantially in halo mass, concentration, redshift, and cross section from the 29 models used to build the emulator. We use the GTF code to evolve haloes A and B deep into core collapse and then compare their properties against the predictions from the emulator. 
\renewcommand{\arraystretch}{1.05}
\begin{table}
\label{tab:emulator_test}
\centering
\renewcommand{\arraystretch}{1.1}
\begin{tabular}{l|ll}
\hline
\multicolumn{3}{l}{Models for emulator validation} \\
\hline
\noalign{\vskip 3pt}
Model         & A & B \\
\hline
$M_{\rm vir}/(\Msunh)$        & $2 \times 10^{13}$  & $3 \times 10^{10}$ \\
$c_{\rm vir}$                 & $17.5$              & $45.5$ \\
$z$                           & $0.2$               & $0.5$ \\
$\sigma_\rmm/(\cmg)$          & $1.0$               & $59.5$ \\
$\hat{\sigma}_\rmm$           & $0.212$             & $8.43$ \\
$(\delta\log \rho)_{\rm max}$ & $0.01897 \; (+4.47\%)$ & $0.01582 \; (-3.58\%)$ \\
$(\delta\log \rho)_{\rm rms}$ & $0.00067 \; (0.154\%)$ & $0.00097 \; (0.224\%)$ \\
$(\delta\log v_c)_{\rm max}$  & $0.00652 \; (-1.49\%)$ & $0.00874 \; (-1.99\%)$ \\
$(\delta\log v_c)_{\rm rms}$  & $0.00007 \; (0.017\%)$ & $0.00015 \; (0.035\%)$ \\
\hline
\end{tabular}
\caption{Parameters of the two models used to test and validate the SIDM core collapse emulator. The final four entries indicate the maximum and rms logarithmic errors in the density and velocity dispersion profiles returned by the emulator.}
\end{table}

Fig.~\ref{fig:emulator_test} shows the evolution of the central core density, core mass, core radius, and core velocity dispersion as a function of time. Blue solid lines show the results obtained directly from the GTF code, while the orange dashed curves show the predictions obtained using the emulator. As is evident, the agreement is excellent. In order to quantify this, the maximum absolute logarithmic residual $(\delta\log X)_{\rm max}$ and the time-weighted RMS logarithmic residual $(\delta\log X)_{\rm rms}$ are indicated (in dex) in each panel. Here, $\delta \log X(t) = \log[X_{\rm emulator}(t)/X_{\rm GTF}(t)]$ with $X$ either core density, core mass, core radius, or the 1D core velocity dispersion.

We also compute the maximum and RMS logarithmic errors for the density and  velocity dispersion profiles. Define
\begin{equation}
\delta \log X_{ij} = \log\left[{X_{\rm emulator}(r_{ij},t_j) \over X_{\rm GTF}(r_{ij},t_j)}\right]
\end{equation}
where $j=0,N$ indicates the time step and $i=1,N_r(j)$ is the number of radial bins at that time\footnote{Recall that in the GTF code this is time step dependent because of the adaptive radial binning.}. Then, the maximum and RMS logarithmic errors are given by
\begin{align}
(\delta\log X)_{\rm max} & = \max_{i,j} \vert\delta \log X_{ij}\vert \\
(\delta\log X)_{\rm rms} & = \left[\sum_{j=0}^{N} w_j\,\frac{1}{N_{r,j}}\sum_{i=1}^{N_{r,j}} \left( \delta\log X_{ij} \right)^2 \right]^{1/2} 
\end{align}
where
\begin{equation}
w_j=
\begin{cases}
\dfrac{t_1-t_0}{2T}, & j=0,\\[6pt]
\dfrac{t_{j+1}-t_{j-1}}{2T}, & 1\leq j\leq N-1,\\[6pt]
\dfrac{t_N-t_{N-1}}{2T}, & j=N,
\end{cases}
\qquad T=t_N-t_0.
\end{equation}
These weights give equal weight per unit physical time, preventing the densely sampled late collapse from dominating merely because of the higher output cadence. Using the 423 (343) radial profiles for Model A (B), we obtain the values for $(\delta\log X)_{\rm max}$ and $(\delta\log X)_{\rm rms}$ indicated in Table~A.1. RMS errors are solidly at the sub-per-cent level for both the density and the velocity dispersion profiles. Hence, we conclude that the emulator returns reliable structural properties of NFW haloes during any stage of their gravothermal collapse.

\section{Computing source magnification}
\label{App:magn}

Here we outline how we compute the total source magnification $\mu(t)$, given by equation~(\ref{magnification}).

Let $I_a(\boldsymbol{s})$ be the source's surface-brightness profile, normalized to unit unlensed flux:
\begin{equation}
\int I_a(\boldsymbol{s}) \, \rmd^2\boldsymbol{s} = 1\,.
\end{equation}
Here $\boldsymbol{s}$ is the displacement from the source centre, and $a$ is an angular size parameter. For the circular Plummer source of equation~(\ref{source}), $a = R_{1/2}$ is the angular half-light radius, and the normalized surface-brightness profile is given by
\begin{equation}
I_a(\boldsymbol{s}) = I_a(s) = {a^2 \over \pi (a^2 + s^2)^2}\,,\qquad s=|\boldsymbol{s}|\,.
\end{equation}

Since gravitational lensing preserves surface brightness, the normalized image brightness at $\boldsymbol{\theta}$ is $I_a[\boldsymbol{\beta}(\boldsymbol{\theta},t) - \boldsymbol{\beta}_0]$. Here $\boldsymbol{\theta}$ and $\boldsymbol{\beta}$ are the positions in the image and source planes, respectively, and $\boldsymbol{\beta}_0$ is the position of the centre of the extended source. Integrating over the image plane gives the total flux magnification
\begin{equation}\label{magn_from_image}
\mu(\boldsymbol{\beta}_0,t) = \int I_a[\boldsymbol{\beta}(\boldsymbol{\theta},t) - \boldsymbol{\beta}_0] \, \rmd^2\boldsymbol{\theta}\,.
\end{equation}
In the case of a spherical lens with zero external shear and a circular Plummer source, the image-plane angular integral can be performed analytically. Consequently, computing $\mu(t)$ requires only a one-dimensional numerical integral.

Starting from the projected physical surface density $\Sigma(R,t)$, we first compute a table of $\{\theta_i,\alpha_i(t)\}$, with $i=1,...,N$, where $\alpha_i = \theta_i \,\overline{\kappa}(<\theta_i)$ with $\overline{\kappa}(<\theta)$ given by equation~(\ref{kappa_aver}). At arbitrary radii, $\alpha(\theta,t)$ is evaluated by linear interpolation of the table values. 

We define 
\begin{equation}\label{udef}
u(\theta,t) = \theta - \alpha(\theta,t).
\end{equation}
and write the image-plane position as $\boldsymbol{\theta}=\theta(\cos\psi,\sin\psi)$. Then, because the deflection is radial and there is no external shear, the lens equation implies that
\begin{equation}
\boldsymbol{\beta}(\boldsymbol{\theta},t) = u(\theta,t)(\cos\psi,\sin\psi).
\end{equation}
Rotational symmetry allows us to choose the coordinate system so that $\boldsymbol{\beta}_0=(b,0)$. The squared distance from the mapped source position to the source centre then becomes
\begin{equation}
\left|\boldsymbol{\beta}(\boldsymbol{\theta},t)
-\boldsymbol{\beta}_0\right|^2 = u^2 + b^2 - 2ub\cos\psi\,.
\end{equation}

Combining, we have that
\begin{align}\label{mubt}
\mu(\boldsymbol{\beta}_0,t) & = \mu(b,t) = \frac{a^2}{\pi}
\int_0^\infty \rmd\theta\,\theta \times \nonumber\\
& \int_0^{2\pi} \frac{\rmd\psi}
{\left[a^2 + u(\theta,t)^2 + b^2 - 2u(\theta,t) b \cos\psi\right]^2}.
\end{align}
Using that
\begin{equation}
\int_0^{2\pi} \frac{\rmd\psi}{(A - B\cos\psi)^2} = \frac{2\pi A}{(A^2 - B^2)^{3/2}}\,
\end{equation}
equation~(\ref{mubt}) reduces to
\begin{equation}\label{mu_int}
\mu(b,t)= 2a^2 \int_0^\infty \frac{[a^2+u(\theta,t)^2+b^2]}{\left( [a^2+C_{-}^{\,2}] [a^2+C_{+}^{\,2}] \right)^{3/2}} \, \theta \, \rmd\theta\,,
\end{equation}
where
\begin{equation}
C_{\pm} \equiv u(\theta,t) \pm b
\end{equation}
This is the integral we evaluate in order to compute the source magnification in the absence of external shear. Importantly, despite the fact that the remaining integral only depends on the angular radius $\theta$, any arcs or other angular structures are properly accounted for through the exact angular integral; it includes all images without identifying them individually.

The integral of equation~(\ref{mu_int}) is evaluated using standard numerical techniques, splitting the $\theta$ domain according to the nodes of the $\{\theta_i,\alpha_i(t)\}$ table. In particular, each segment of this deflection table is integrated separately using adaptive Simpson quadrature and further subdivided if it crosses the radial or tangential critical radii. We assess numerical convergence by repeating representative calculations with finer radial grids for the density and deflection profiles until the resulting magnifications change by less than $10^{-3}$ in relative terms.

For non-zero external shear, the halo deflection $\alpha(\theta,t)$ is unchanged, but the lens equation becomes
\begin{equation}
\boldsymbol{\beta}(\theta,\psi,t) = [u(\theta,t) - \theta \, \boldsymbol{\Gamma}] \, (\cos\psi,\sin\psi)\,,
\end{equation}
where $\boldsymbol{\Gamma}$ is the external-shear matrix defined in equation~(\ref{shear}). The total magnification is therefore
\begin{equation}\label{mu_shear_int}
\mu(\boldsymbol{\beta}_0,t) = \frac{a^2}{\pi} \int_0^{2\pi} \rmd\psi \int_0^\infty \frac{\theta \, \rmd\theta} {\left[a^2+\left|\boldsymbol{\beta}(\theta,\psi,t) - \boldsymbol{\beta}_0\right|^2 \right]^2}\,.
\end{equation}
Since the magnification depends on the orientation of the shear relative to the source offset, the $\psi$-integration can no longer be done analytically. In order to compute $\mu(\boldsymbol{\beta}_0,t)$, we first perform the radial $\theta$-integration using a numerical technique similar to that described above, and evaluate the $\psi$-integration using uniformly spaced quadrature, doubling the number of angles until convergence. 

Note that no PSF convolution is taken into account when computing source magnifications. PSF convolution is only used to render the images in Fig.~\ref{fig:image_sequence}. Since it conserves total flux, the magnification of the PSF convolved images is identical to that in the absence of convolution.

\subsection{Aperture-averaged magnification}
\label{App:aperture_averaging}

In Section~\ref{sec:sourcepos} we compute the {\it average} magnification $\langle B(t) \rangle$, averaged over a circular aperture of radius $\beta_{\rm max} = 0.25 \arcsec$. Substituting equation~(\ref{magn_from_image}) into equation~(\ref{aver_boost}) and interchanging the order of integration, we have that
\begin{equation}
\langle B(t) \rangle = {1 \over \pi \beta_{\rm max}^2} \int K\left[\boldsymbol{\beta}(\boldsymbol{\theta},t)\right] \, \rmd^2\boldsymbol{\theta}
\end{equation}
where we define the kernel
\begin{equation}\label{kernel}
K(\boldsymbol{u}) = \int_{|\boldsymbol{\beta}_0| \leq \beta_{\rm max}} {I_a(\boldsymbol{u}-\boldsymbol{\beta}_0) \over \mu_0(\boldsymbol{\beta}_0)} \, \rmd^2\boldsymbol{\beta}_0
\end{equation}
which incorporates both the source profile and the source-position-dependent NFW normalization. 

When the shear is non-zero, $K(\boldsymbol{u})$ is a two dimensional function. Rotating the coordinate axes so that the shear matrix is diagonal,
\begin{equation}
\boldsymbol{\Gamma} =
\begin{pmatrix}
\gamma_{\rm ext} & 0 \\
0 & -\gamma_{\rm ext}
\end{pmatrix}
\end{equation}
the lens equation becomes
\begin{align}
\beta_x &= [(1-\gamma_{\rm ext})\theta - \alpha(\theta,t)]\, \cos\chi\,, \nonumber \\
\beta_y &= [(1+\gamma_{\rm ext})\theta - \alpha(\theta,t)]\, \sin\chi\,.
\end{align}
Using that the system in this rotated frame has reflection symmetry about both shear axes, we only need to integrate over one quadrant, i.e.,
\begin{equation}\label{Baver_quadrant}
\langle B(t) \rangle = {4 \over \pi \beta_{\rm max}^2} \int_0^{\infty} \theta \, \rmd\theta \int_0^{\pi/2} K\left[\beta_x(\theta,\chi,t), \beta_y(\theta,\chi,t) \right] \, \rmd\chi
\end{equation}
We first compute $\mu_0(b,\varphi)$ on a source-position grid covering one quadrant ($0 \leq b \leq \beta_{\rm max}$ and $0 \leq \varphi \leq \pi/2$), which we use to precompute the two-dimensional kernel $K$ by treating equation~(\ref{kernel}) as a convolution of $I_a$ and the function 
\begin{equation}
W(\boldsymbol{\beta}_0) = 
\begin{cases}
1/\mu_0(\boldsymbol{\beta}_0), & |\boldsymbol{\beta}_0| < \beta_{\rm max}\,,\\
0,                             & \text{otherwise}.
\end{cases}
\end{equation}
which we evaluate using an FFT. Using bilinear interpolation of the resulting two-dimensional kernel $K$, equation~(\ref{Baver_quadrant}) is subsequently integrated using standard quadrature techniques, interpolating $\alpha(\theta,t)$ linearly between the nodes of the $\{\theta_i,\alpha_i(t)\}$ table in order to compute $\beta_x$ and $\beta_y$.

\subsection{Magnification due to separate images}
\label{App:image_magn}

When evaluating the strong lensing efficacy parameter $\Gamma_2$ of equation~(\ref{Gamma2}), we need to compute the magnifications of the two brightest images, $\mu_1$ and $\mu_2$. As mentioned in Section~\ref{sec:multi_eff}, this is done using an intrinsic image-branch criterion.

Suppose an extended source overlaps a radial caustic. Parts of the source then produce images on both sides of the corresponding radial critical curve. The image-branch method described below assigns the brightness on each side to its respective branch, even if the two bright regions touch and appear as one connected arc. Likewise, a nearly centred source can produce a ring-like image. The inner and outer sides of the tangential critical curve belong to different branches, whether or not an observer can resolve or deblend them.

In the case of a spherical lens with zero external shear, the two Jacobian eigenvalues are
\begin{equation}
\lambda_\rmt = \frac{u(\theta)}{\theta},
\qquad
\lambda_\rmr = \frac{\rmd u}{\rmd\theta}.
\end{equation}
where $u(\theta,t)$ is given by equation~(\ref{udef}). Thus, tangential and radial critical curves occur where $u=0$ and $\rmd u/\rmd\theta=0$, respectively. For the familiar case with one radial critical radius $\theta_\rmr$ and one tangential critical radius $\theta_\rmE$, there are three branches: (i) $0 < \theta < \theta_\rmr$, (ii) $\theta_\rmr < \theta < \theta_\rmE$, and (iii) $\theta > \theta_\rmE$. If the mapping has additional radial extrema inside the Einstein radius, additional branches are introduced along similar lines.

For our fiducial Plummer source, the magnification of image $j$ belonging to the branch that occupies the radial interval $\theta_{j,-} < \theta < \theta_{j,+}$ is given by equation~(\ref{mu_int}), but with the integration interval $[0,\infty]$ replaced by $[\theta_{j,-},\theta_{j,+}]$. After computing the various branch fluxes, we sort them in descending order and use $\mu_1$ and $\mu_2$ to evaluate $\Gamma_2$

\section{Lensing due to SIDM cores}
\label{App:cores}

Consider a region with nearly constant surface density, $\Sigma_\rmc$. Its convergence $\kappa_\rmc = \Sigma_\rmc/\Sigma_{\rm crit}$ is also roughly constant. Hence, using equation~(\ref{alphatheta}) we have that
\begin{equation}
\alpha(\theta) = \kappa_\rmc \, \theta
\end{equation}
Using the lens equation~(\ref{lensequation}) this implies that
\begin{equation}
\beta = \theta \, (1-\kappa_\rmc)
\end{equation}
Hence, a large area in the image plane will map to a small area in the source plane if $\kappa_\rmc \sim 1$. Note also that, since $\bar{\kappa}_\rmc = \kappa_\rmc$, we have that $\lambda_\rmt = \lambda_\rmr = 1-\kappa_\rmc$, which implies a magnification
\begin{equation}\label{core_magn}
\mu = (1-\kappa_\rmc)^{-2}
\end{equation}
which indicates that such mapping results in large magnification, as expected.

This demonstrates that high magnification requires a large image-plane area to map onto bright parts of the source. It does not require a large central surface density. In fact, as is evident from equation~(\ref{core_magn}), increasing $\Sigma_\rmc$ to highly supercritical values will result in reduced magnification. 

Consider an idealized lens with constant projected convergence $\kappa_\rmc$ inside an angular radius $\theta_\rmc = r_\rmc/D_\rmL$, and zero surface density outside. Then, the contribution to magnification from image positions located inside the core region is
\begin{equation}\label{mu_in}
\mu_{\rm in}
= 2 \pi \int_0^{\theta_\rmc} I_a \, \left(|1-\kappa_\rmc|\theta \right) \, \theta\, \rmd\theta\,.
\end{equation}

Let the source be our fiducial, circular Plummer source located at the origin, so that
\begin{equation}
I_a(\beta) = \frac{a^2}{\pi (a^2 + \beta^2)^2}\,,
\qquad a=R_{1/2}\,.
\end{equation}
Substitution in equation~(\ref{mu_in}) yields
\begin{equation}
\mu_{\rm in} = \frac{\theta_\rmc^2}{a^2 + (1-\kappa_\rmc)^2 \theta_\rmc^2}\,.
\end{equation}
which captures the competition between core size and core density. When $\Sigma_\rmc = \Sigma_{\rm crit}$, we have that
\begin{equation}
\mu_{\rm in} = \left( {\theta_\rmc \over a} \right)^2\,.
\end{equation}
Hence, for a fixed source size, the magnification scales as $\theta^2_\rmc$. Thus, a constant core yields high magnification when its surface density is close to critical, and the actual magnification it achieves scales with the radius of the core squared. In the case of an SIDM halo, once the core density becomes supercritical and the core radius continues to decrease, the magnification drops precipitously, as indeed witnessed in the numerous cases presented throughout the main text.  Note that the expressions above only describe the contribution from image positions inside the core. The total magnification also includes images outside the core, which can contribute substantially even for a compact source. The decline of the total magnification in our SIDM models is established by the full calculations in the main text.

\label{lastpage}
\end{document}